\documentclass[%
 reprint,
 amsmath,amssymb,
 aps,
prc,
]{revtex4-1}
\usepackage{lipsum}
\usepackage{graphicx}
\usepackage{dcolumn}
\usepackage{bm}
\usepackage{amssymb}
\RequirePackage{amsmath}
\RequirePackage{amsfonts}
\RequirePackage{graphicx}
\usepackage{fontenc}
\usepackage{longtable}
\usepackage{epstopdf}
\DeclareGraphicsExtensions{.eps}
\begin{document}

\preprint{APS/123-QED}

\title{Evolution of E2 transition strength in deformed hafnium isotopes from
new measurements on $^{172}$Hf, $^{174}$Hf, and $^{176}$Hf}
\author{M. Rudigier$^1$}
\email[email: ]{rudigier@ikp.uni-koeln.de}

\author{K. Nomura$^{2,1}$}

\author{M. Dannhoff$^1$}
\author{R-B. Gerst$^1$}
\author{\mbox{J. Jolie$^1$}}
\author{N. Saed-Samii$^1$}
\author{S. Stegemann$^1$}
\author{J-M. R\'egis$^1$}

\author{L. M. Robledo$^3$}
\author{R. Rodr\'iguez-Guzm\'an$^4$}

\author{A. Blazhev$^1$}
\author{Ch. Fransen$^1$}
\author{N. Warr$^1$}
\author{K.O. Zell$^1$}

\affiliation{$^1$Institut f\"ur Kernphysik, Universit\"at zu K\"oln,
Z\"ulpicher Str. 77, D-50937 K\"oln, Germany}
\affiliation{$^2$ Grand Acc\'el\'erateur National d'Ions Lourds,
CEA/DSM-CNRS/IN2P3, B.P.55027, F-14076 Caen Cedex 5, France}
\affiliation{$^3$Departmento de F\'isica Te\'orica, Universidad Aut\'onoma de
Madrid, E-28049 Madrid, Spain}
\affiliation{$^4$Physics Department, Kuwait University, Kuwait 13060, Kuwait}


\date{\today}

\begin{abstract}

\begin{description}

\item[Background] The available data for E2 transition strengths in the region
between neutron deficient hafnium and platinum isotopes is far from complete.
More and precise data are needed to enhance the picture of structure evolution
in this region and to test state-of-the-art nuclear models. In a simple model,
the maximum collectivity is expected at the middle
of the major shell. However, for actual nuclei particularly in heavy
mass regions, which should be highly complex, this picture may no
longer be the case, and one should use a more realistic nuclear
structure model. We address this point by studying the spectroscopy
of Hf as a representative case.

\item[Purpose] To re-measure the $2^+_1$ half-lives of
$^{172,174,176}$Hf, for which there is some disagreement
in the literature. The main goal is to measure, for the first
	   time, the half-lives of
higher lying states of the rotational band. 
The new results are compared to a theoretical
	   calculation for absolute transition strengths.

\item[Method] The half-lives were measured using $\gamma$-$\gamma$ and 
conversion
electron-$\gamma$ delayed coincidences with the fast timing method. For the
determination of half-lives in the picosecond region the generalized centroid
difference method was applied. 
For the theoretical calculation of the spectroscopic properties, the
	   interacting boson model is employed, whose Hamiltonian is
	   determined based on microscopic energy-density functional 
calculations. 

\item[Results] The measured $2^+_1$ half-lives disagree with results from
earlier $\gamma$-$\gamma$ fast timing measurements, but are in agreement with 
data
from Coulomb excitation experiments and other methods. Half-lives of the
$4^+_1$ and $6^+_1$ states were measured, as well as a lower limit for the
$8^+_1$ states.

\item[Conclusions]

This work shows the importance of a mass dependent effective boson 
charge in the interacting boson model for the description of E2 transition 
rates 
in chains of nuclei. It encourages further studies of the microscopic origin of 
this mass dependence. New experimental values on transition rates in nuclei 
from 
neighboring isotopic chains could support these studies.

\end{description}
\end{abstract}

\maketitle

\section{Introduction}\label{sec:introduction}

The absolute strength of E2 transitions between low-lying states of
even-even nuclei is an important observable to test nuclear models that
describe collective phenomena. Well deformed even-even nuclei exhibit large
quadrupole transition strengths \mbox{$B(E2;2^+_1 \rightarrow 0^+_1)$} 
\cite{BM}.
In a simple picture this strength increases smoothly
for increasing number of valence nucleons or holes along an isotopic or 
isotonic chain as collectivity increases. Assuming symmetry of particles and 
holes this model yields a maximum at mid-shell \cite{CasBook}.
Microscopic effects can break the particle-hole symmetry and lead to a 
different picture.

Recent measurements on tungsten and osmium isotopes
\mbox{\cite{PhysRevC.86.057303, Williams12, Rudigier201089}} showed that for a
given isotopic chain the maximum
of the absolute strength \mbox{$B(E2; 2^+_1 \rightarrow 0^+_1)$} is not found
at mid-shell but at lower neutron number. Furthermore there seems to be a sudden
increase at N = 98, which is not expected for a collective observable.
The overall picture of available data on B(E2)s is rather erratic for these and
neighboring nuclei, which makes a comparison to nuclear structure models
difficult. In many cases important data are missing or there exist disagreeing
results from experiments using different methods.
For a better understanding of nuclear structure evolution in this region it is
important to have a more complete picture of absolute transition strengths for
the isotopes with $70\le Z\le78$. 

In this paper, new $B$(E2) values for hafnium
isotopes (Z=72) are presented. They were measured using the method of delayed
coincidences with LaBr$_3$(Ce) detectors and an Orange conversion electron
spectrometer. The results improve the data situation and enable us to
test current model predictions of both absolute and relative transition strengths.

On the theoretical side, a quantitative and detailed description of
spectroscopic properties for heavy nuclei has been provided by fully-microscopic many-body
theories, which include the large-scale shell-model \cite{brown01,taka01,caurier05} and the self-consistent
mean-field \cite{Ben03rev,Nik11rev,erler11}, or the energy density 
functional (EDF), approaches with a suitably chosen effective nucleon-nucleon 
interaction. 

In the case of strongly deformed heavy nuclei, like the ones considered here, 
the dimension of the shell-model configuration space becomes exceedingly large
requiring an appropriate truncation scheme to reduce the computational cost
while keeping the essential features of low-lying collective states. 
Along this line, one could assume that the nuclear many-body problem is 
approximated by a system of interacting bosons: the interacting boson model (IBM) \cite{IBM}. 
The essence of the model is the  association of the collective nucleon pairs
relevant for the low-energy states, 
e.g., with spin $J=0^+$ ($S$), $2^+$ ($D$), \ldots, to the equivalent bosonic degrees of
freedom ($s$, $d$, \ldots) by means of a well defined mapping procedure.
The physical states and their decay properties are 
obtained from the calculations in the boson space \cite{taka78,OAI}. 

In recent years, the IBM has been used with considerable success in
spectroscopic studies \cite{mccutchan04,MCC05,zerguine08,zerguine12,Kot12} 
of a large set of nuclei with proton and neutron
numbers in the range $Z=50-82$ and  $N=82-126$, respectively. 
However, those studies are rather phenomenological because the
parameters of the IBM Hamiltonian are determined by a fit to known
experimental data.

On the other hand, the mean-field framework has also been
successfully applied to the study of nuclear structure phenomena \cite{Ben03rev}. 
Both in the non-relativistic
\cite{Sky59,Gogny} and
relativistic \cite{vretenar05,Nik11rev} regimes,  
it is possible to  obtain a global and reasonable description of
the ground-state properties and
collective excitations of all nuclei across the nuclear landscape with
a single parametrization of the corresponding EDF.
However, to describe in detail spectroscopic properties one
needs to go beyond the mean-field level to restore symmetries broken in
the mean field approximation as well as to take into account fluctuations with
collective coordinates. 
Much effort has been devoted to increase the feasibility of such calculations,
mostly in the framework of the pure generator coordinate method 
\cite{rayner02,Ben03rev,nik07,bender08,robledo11} or its approximations \cite{delaroche10,nom11coll}.
The results confirm the usefulness and  reliability of these EDF-based
approaches for the study of nuclear spectroscopy. 

To investigate the spectroscopic properties of the 
neutron-deficient hafnium isotopes, we use the procedure of Ref.~\cite{Nom08},
that determines the Hamiltonian of the IBM from EDF-based mean-field results. 
The idea behind this method is to map the deformation energy surface
resulting from a set of the constrained mean field calculations onto the
equivalent energy surface for the system of interacting bosons, that is, onto the
expectation value of the IBM Hamiltonian in the boson condensate state \cite{GK}. 
The parameters of the IBM Hamiltonian determined by this procedure do not require
any additional adjustment to experimental data.
The resulting Hamiltonian is used to calculate energy spectra and transition rates. 
So far, the predictive power of the method has been verified in all the possible regimes of
low-energy quadrupole collective states: spherical 
vibrational \cite{Nom10}, $\gamma$-soft \cite{Nom10,nom12tri} and
well-deformed rotational \cite{Nom11} nuclei. 


Given the predictive power of our procedure it is interesting to check whether the IBM
Hamiltonian determined from the microscopic mean-field calculation can 
explain the absolute transition rates for the neutron deficient hafnium isotopes. 
The microscopic input used is the Gogny energy density functional with the
recent parametrization D1M \cite{D1M}, which has been shown
\cite{rayner10pt,rayner10odd-3,rayner12} to have a similar level of 
predictive power in the description of nuclear structure phenomena as
the standard parametrization D1S \cite{D1S}. 
However, and to confirm the robustness of our
results, we also discuss the 
results obtained with the more traditional Gogny D1S parametrization. 

This paper is organized as follows. In Sec.~\ref{sec:experiment} we
describe the experimental procedure and the data analysis. We then
present the results of the experiment in Sec.~\ref{sec:results}. 
In Sec.~\ref{sec:calculations}, we compare the present experimental results with the
predictions by the IBM-2 calculation combined with the self-consistent
mean-field method using the Gogny energy-density functional. 
A conclusion is given in Sec.~\ref{sec:conclusion}. 
Finally, in Appendix \ref{sec:mapping}, the theoretical procedure is
described in detail.

\section{Experimental procedure and data analysis}\label{sec:experiment}
\begin{figure}
\includegraphics[width=.9\columnwidth]{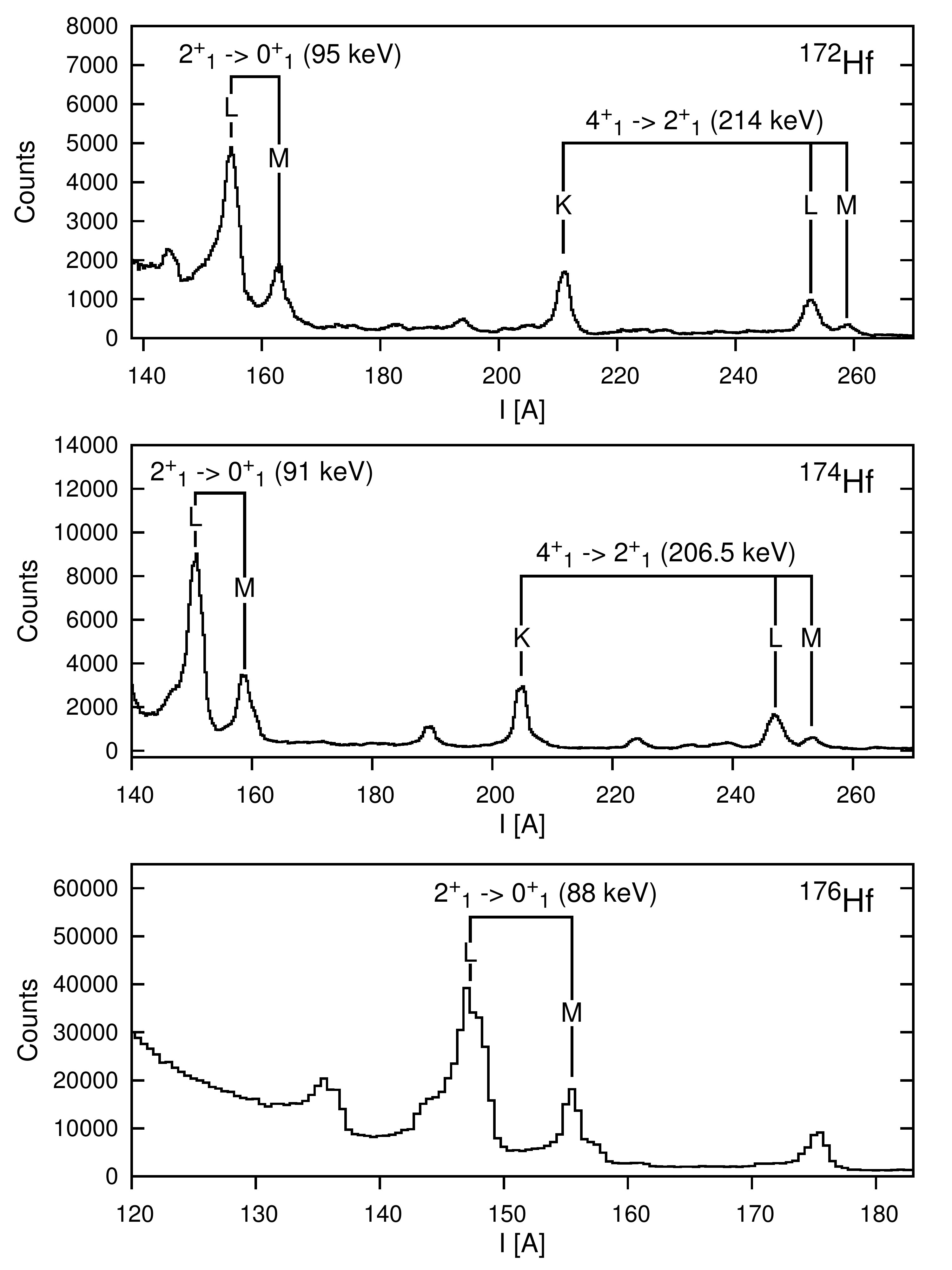}
\caption{Conversion electron momentum spectra of the three investigated nuclei 
($I$ is the current applied to the magnetic spectrometer, see text for 
more details).
Different conversion lines from the same nuclear transition are
marked accordingly. The step intervals during the scans were
0.3 A ($^{172}$Hf and $^{174}$Hf) and 0.5 A ($^{176}$Hf).}\label{fig:ceScans}
\end{figure}

The experiments were performed at the Institut f\"ur Kernphysik of the
University of Cologne. The Cologne FN tandem accelerator delivered an
$\alpha$ beam to induce the reactions \mbox{$^{170}$Yb($\alpha$,
2n)$^{172}$Hf} and \mbox{$^{172}$Yb($\alpha$, 2n)$^{174}$Hf} at a beam energy
of 27 MeV, and \mbox{$^{174}$Yb($\alpha$, 2n)$^{176}$Hf} at a beam energy of
26 MeV. The target thickness was 0.40 mg/cm$^2$, 0.40 mg/cm$^2$, and 0.42
mg/cm$^2$, respectively. Thin targets were chosen in order to minimize
energy straggling and absorption of emitted internal conversion electrons
(ce). These were measured with the Cologne Orange spectrometer, a
toroidal magnetic spectrometer. The electrons are measured with a fast
plastic scintillator detector at the exit slit of the spectrometer. See
reference \cite{Reg09} for more details on this instrument. A scan with
different magnetic fields, i.e. different electric currents, yields an
electron momentum spectrum with the different conversion lines corresponding
to each nuclear transition. Figure~\ref{fig:ceScans} shows ce spectra
measured for the three investigated reactions. The exponential background at
low energies is due to $\delta$ electrons, which are produced by the beam
ions traversing the target. Note that these electrons are not correlated in
time with the decay radiation emitted after a nuclear reaction.

\begin{figure}
\includegraphics[width=\columnwidth]{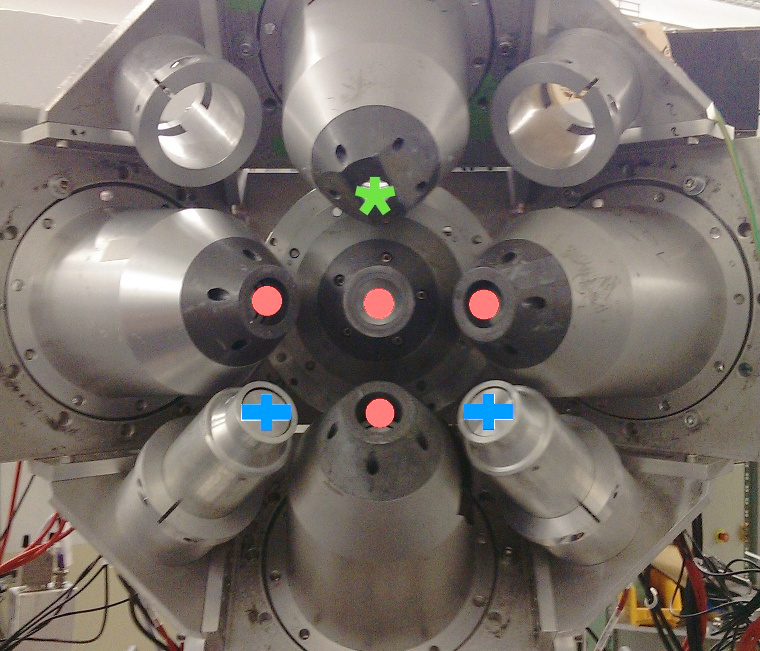}
\caption{(Color online) Photograph of the $\gamma$ detector array in
maintenance position. The position of the four BGO shielded LaBr detectors
(red circles) and the two unshielded LaBr detectors (blue crosses) can be seen.
Only the BGO shield is mounted where the germanium detector was
positioned during the experiment (green star). The view is from the target
position, the beam direction is from right to left.
}\label{fig:Array}
\end{figure}

\begin{figure}
\includegraphics[width=.9\columnwidth]{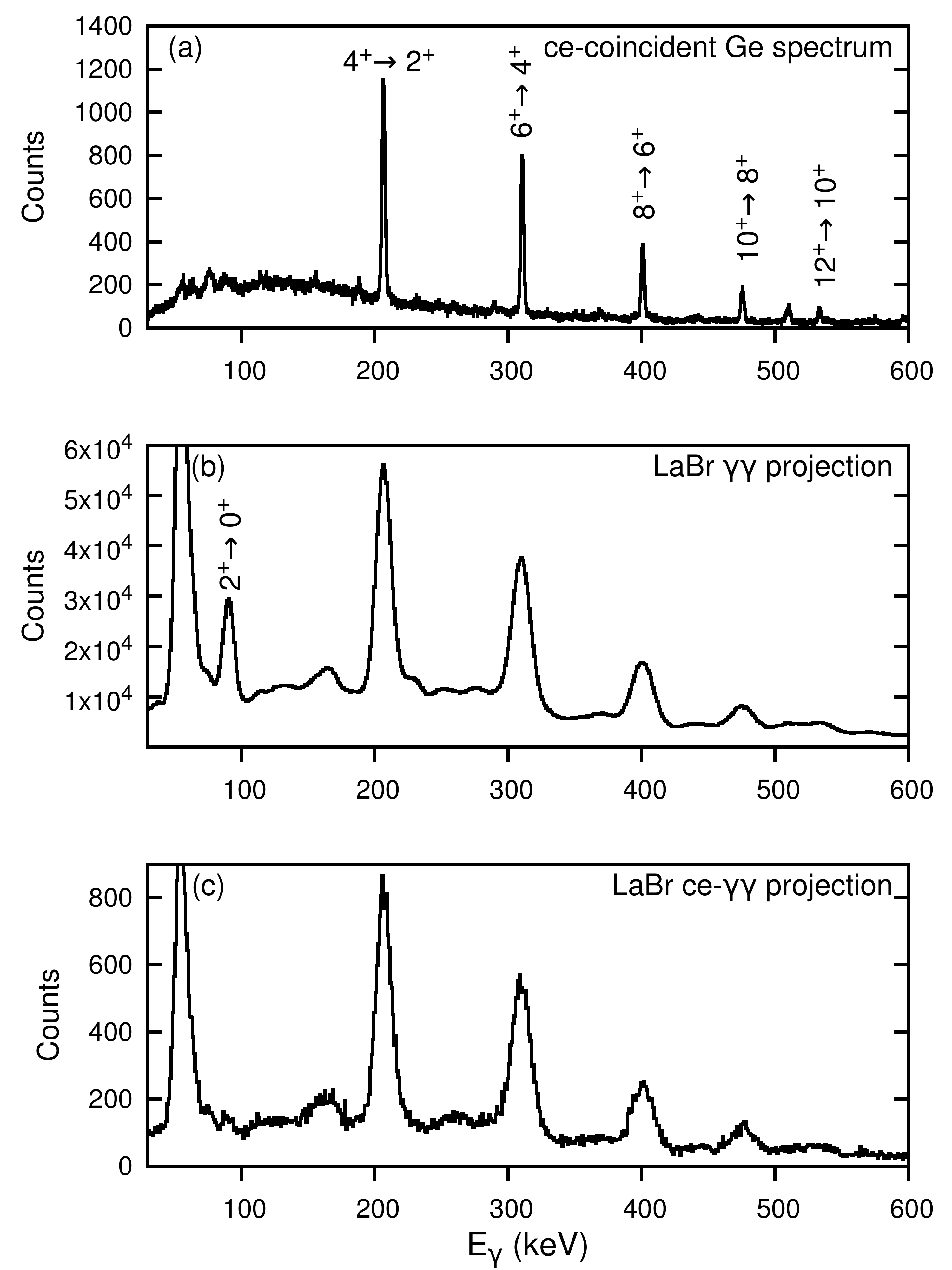}
\caption{$\gamma$-ray spectra of $^{174}$Hf from one experimental run
with different coincidence conditions.
(a) A spectrum measured with the Ge detector
    in coincidence with electrons corresponding to the L-conversion of
    the 2$^+_1$ $\rightarrow$ 0$^+_1$ transition.
    The Ge detector was shielded with lead and copper to suppress X rays.
(b) LaBr $\gamma$-$\gamma$ projection without electron coincidence condition.
(c) LaBr $\gamma$-$\gamma$ projection with the same electron coincidence 
condition
    as in (a).
}\label{fig:Y2plus}  
\end{figure}

The $\gamma$-rays were measured using a small array of six LaBr$_3$(Ce)
(hereafter called LaBr) scintillator detectors and one high purity germanium
detector (HPGe) (see Fig. \ref{fig:Array}) which were mounted perpendicular
to the beam next to the target position of the Orange spectrometer. The  
LaBr-crystals were
cylindrical and 1.5'' $\times$ 1.5'' in size. Four of the LaBr detectors were
equipped with bismuth germanate (BGO) Compton shields and conical lead
collimators to provide active Compton suppression and passive shielding.
The reduction of Compton events generated in the LaBr
crystal and the active suppression of stray $\gamma$ rays produced by 
primary Compton scattering in the experimental surrounding is important
in delayed coincidence timing measurements. Such
background events, including inter-detector Compton
scattering (cross-talk events), are time correlated
and the determination of their timing contribution poses a
major source of uncertainty of the final result.
The two remaining LaBr detectors were unshielded. The HPGe detector was 
installed perpendicular
to the beam axis. Its main purpose during the experiment was to monitor the
reaction, taking advantage of its energy resolution, which is superior to that
of the LaBr detectors (see Figure~\ref{fig:Y2plus}(a)). In the analysis the
HPGe data was used to confirm the level schemes and to identify contaminations
which could otherwise possibly be overlooked in the LaBr spectra. This was
especially important for the $^{172}$Hf data, as $^{172}$Hf is $\beta$
unstable, albeit with a long half-life of 1.87 years \cite{NNDC172}.

Lifetimes were deduced with the method of delayed coincidences. The time
difference between two signals was measured using time to amplitude
converters (TACs) arranged in a fast timing circuit as described in
\cite{Reg13}. The data were recorded triggerless in a list mode format and
analyzed offline. This way it was possible to sort the data with a double
(ce$\gamma$ and $\gamma\gamma$) as well as with a triple coincidence condition
(ce$\gamma\gamma$).
LaBr $\gamma\gamma$ projection spectra for both cases are shown in Fig.~\ref{fig:Y2plus}(b,c).

\subsection{Half-life of $2^+_1$ states}
\begin{figure}
\includegraphics[width=.9\columnwidth]{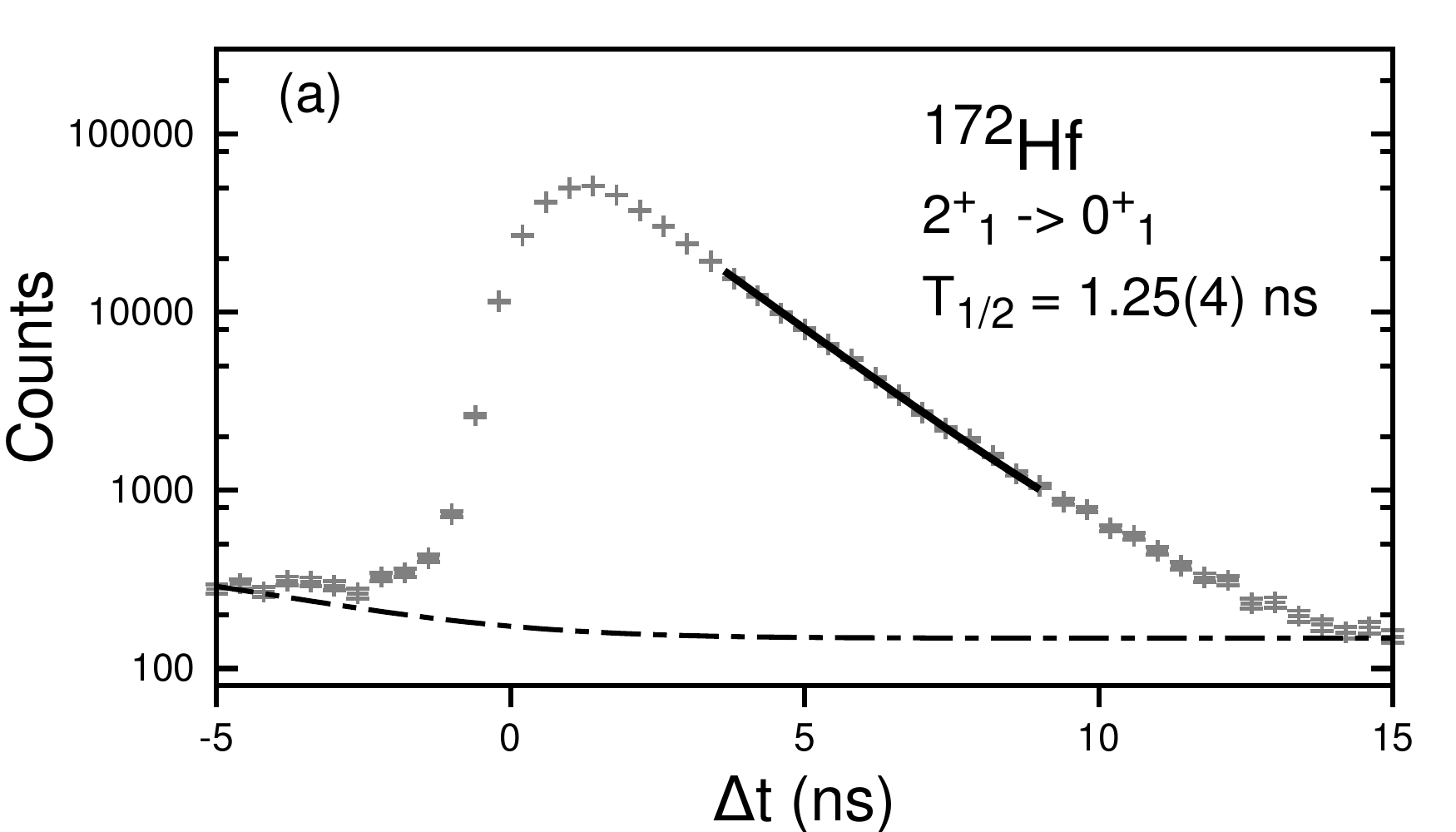}
\includegraphics[width=.9\columnwidth]{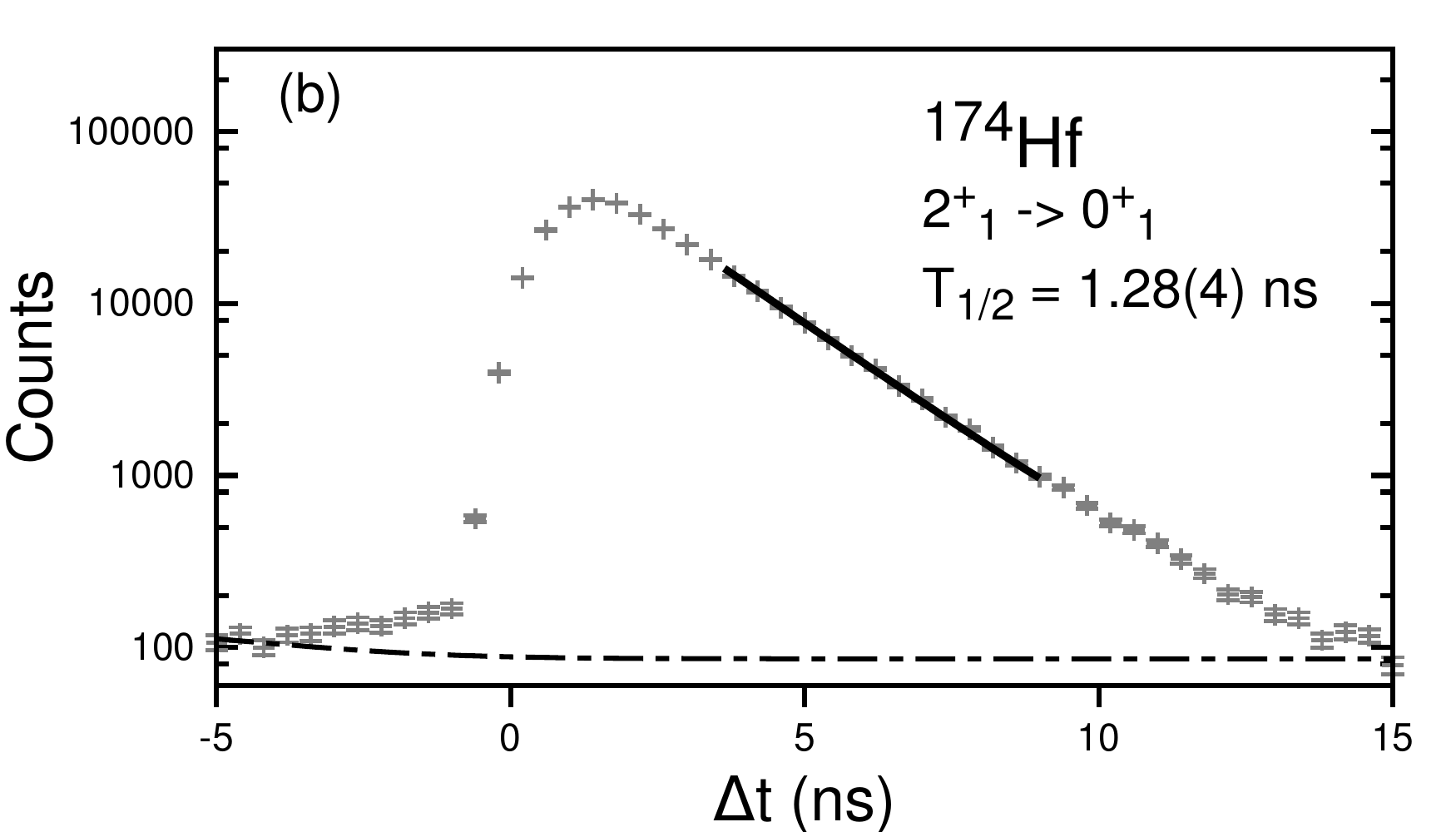}
\includegraphics[width=.9\columnwidth]{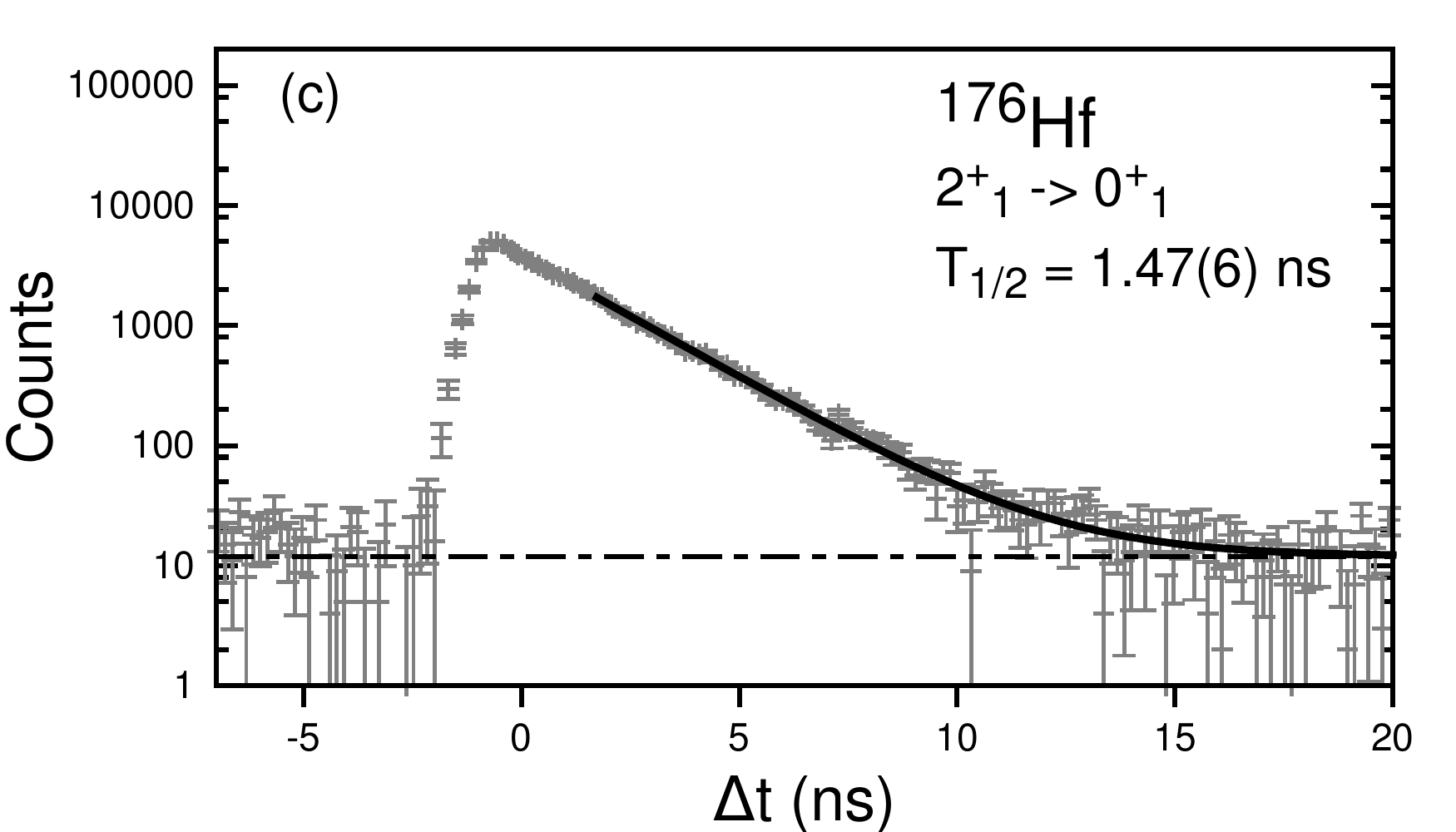}
\caption{ce-$\gamma$ (a,b) and $\gamma$-$\gamma$ (c) delayed coincidence time
spectra of the cascade \mbox{$4^+_1 \rightarrow 2^+_1 \rightarrow 0^+_1$}.
Compton background has been subtracted to yield spectrum (c). The half-life was
determined by fitting the slope. The best fit of an exponential function is
shown as a solid line. The random background, shown as a dashed line, was 
determined separately.}\label{fig:2plus_timespecs}
\end{figure}

The half-life of the 2$^+_1$ state of $^{172}$Hf and $^{174}$Hf
was determined using ce-$\gamma$ coincidences. For this purpose the Orange
spectrometer was set to the electric current corresponding to the
L-conversion line of the \mbox{2$^+_1$ $\rightarrow$ 0$^+_1$} transition.
The L-conversion line was preferred over the K-conversion line because the
latter lies at 23 keV, where it is buried in the $\delta$-electron
background, which increases exponentially towards lower energies.
Furthermore, the K-conversion coefficient $\alpha_\mathrm{K}$ is smaller
than $\alpha_\mathrm{L}$ for the \mbox{2$^+_1$ $\rightarrow$ 0$^+_1$}
transition in the investigated nuclei. E.g. \mbox{$\alpha$ = 5.77},
\mbox{$\alpha_\mathrm{K}$ = 1.18},
\mbox{$\alpha_\mathrm{L}$ = 3.49} for an 88.35 keV E2 transition in $^{176}$Hf.
A LaBr energy gate was then set on the \mbox{4$^+_1$ $\rightarrow$ 2$^+_1$}
transition to produce the TAC spectra shown in Figure~\ref{fig:2plus_timespecs}
(a,b). Only the Compton suppressed LaBr detectors
were used. A fit of the slope yields the half life of the first excited 2$^+$
state. Several different parameterizations of the random background were tried. 
All
variations were consistent within the uncertainty. The results are shown in
Table \ref{tab:results_172Hf} - \ref{tab:results_176Hf}.

In the $^{176}$Hf experiment we encountered a problem with the TAC that
was started by the electron detector which was not noticed until after the
beam time. It was not possible to extract reliable ce-$\gamma$ time spectra
and the $\gamma$-$\gamma$ approach was therefore used in this case.
The \mbox{$2^+_1 \rightarrow 0^+_1$} transition is highly converted and at very
low energy and the $\gamma$ line in the LaBr spectra is therefore weak and
on top of a lot of time-correlated background. The background subtraction,
which yields the spectrum shown in Figure~\ref{fig:2plus_timespecs}
(c), increases the uncertainty dramatically. The result agrees well with the
value from Coulomb excitation measurements given in the nuclear data sheets
\cite{NNDC176}.

\subsection{Half-life of higher lying yrast states}

\begin{figure}
\includegraphics[width=\columnwidth]{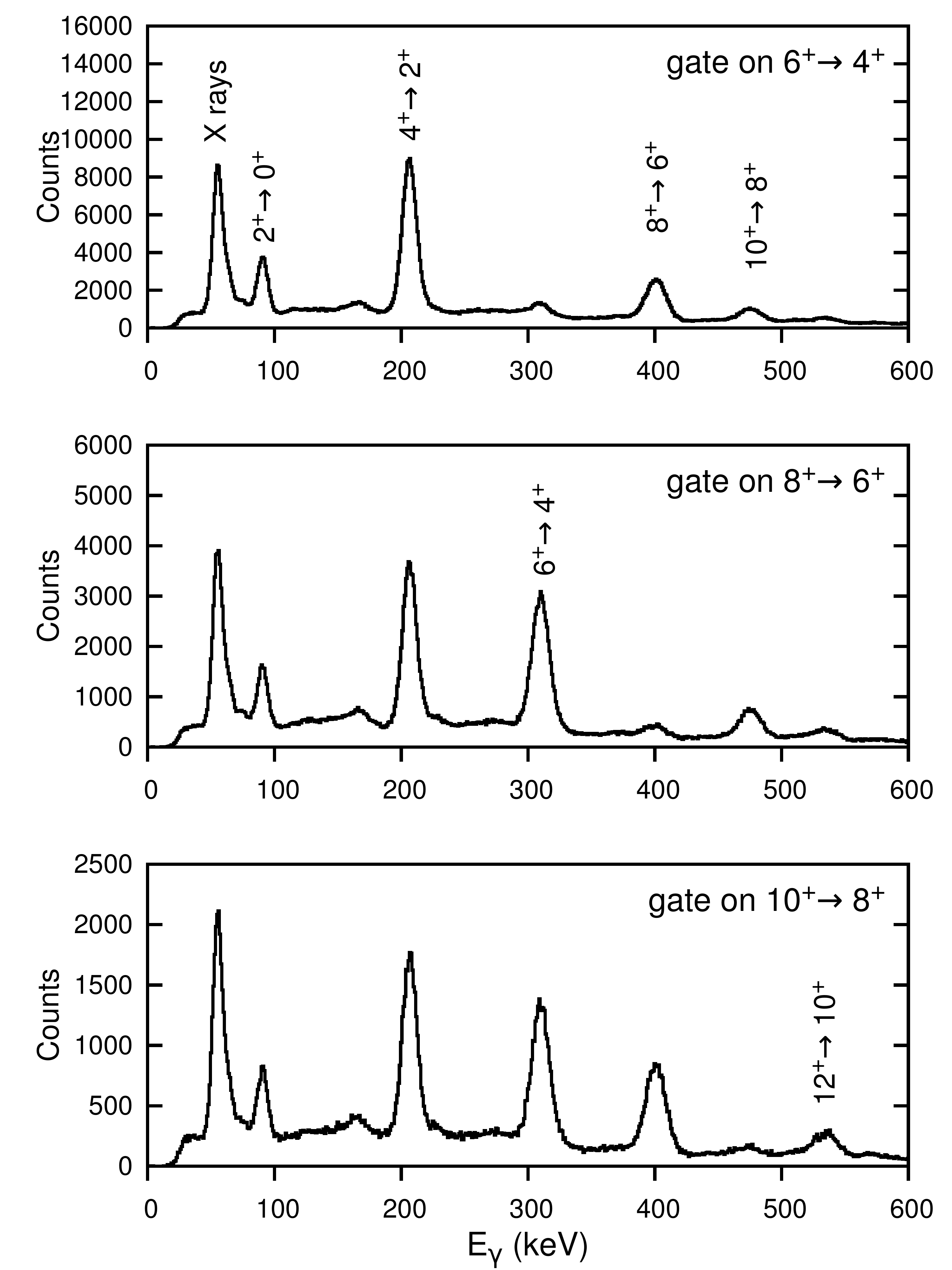}
\caption{Examples of gated LaBr $\gamma\gamma$ coincidence spectra
for $^{174}$Hf. The transition which was gated on
is indicated in the top-right corner of each panel.
No coincident electron was demanded.}\label{fig:Yspecs}
\end{figure}

\begin{figure}
\includegraphics[width=\columnwidth]{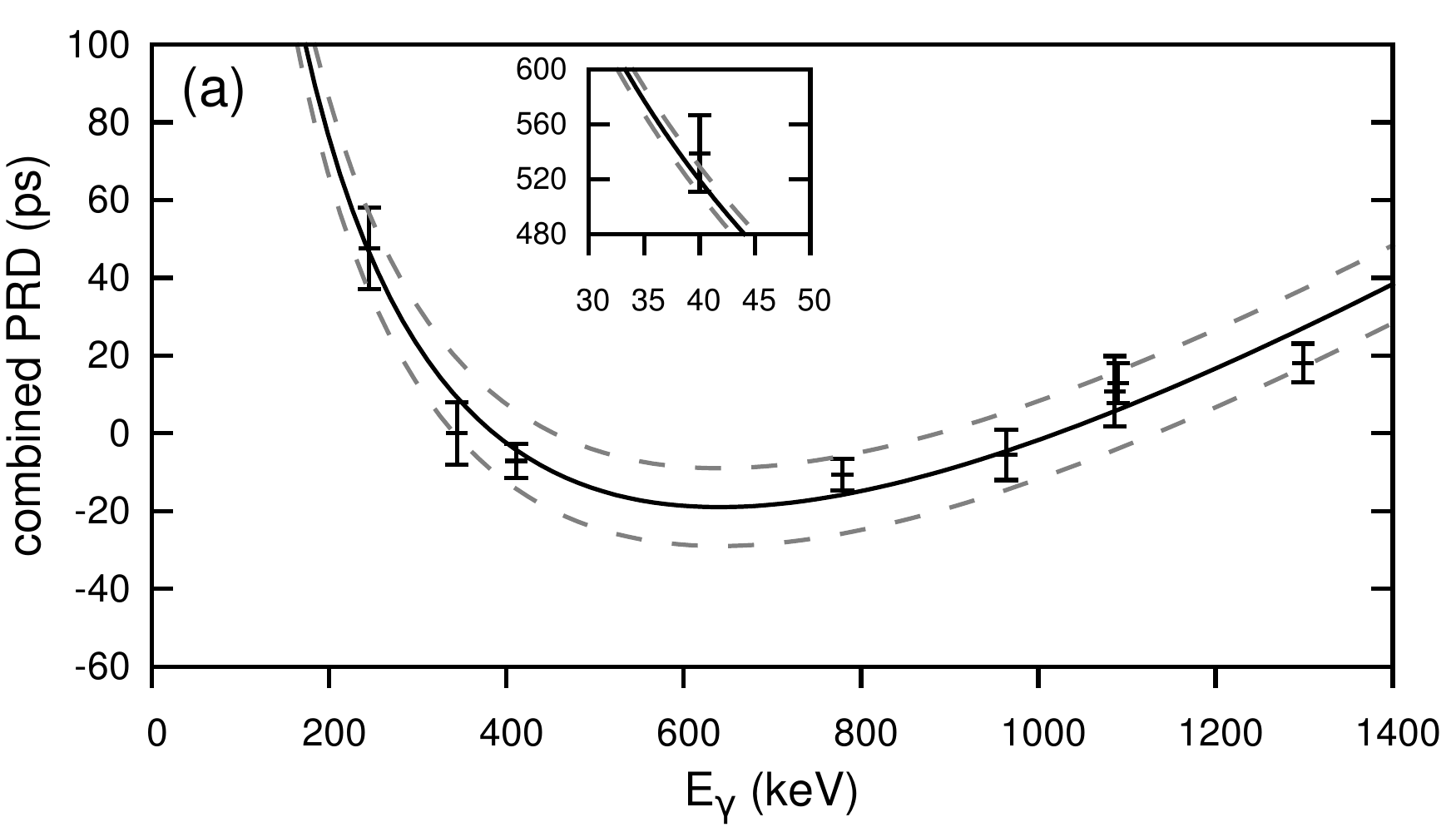}
\includegraphics[width=\columnwidth]{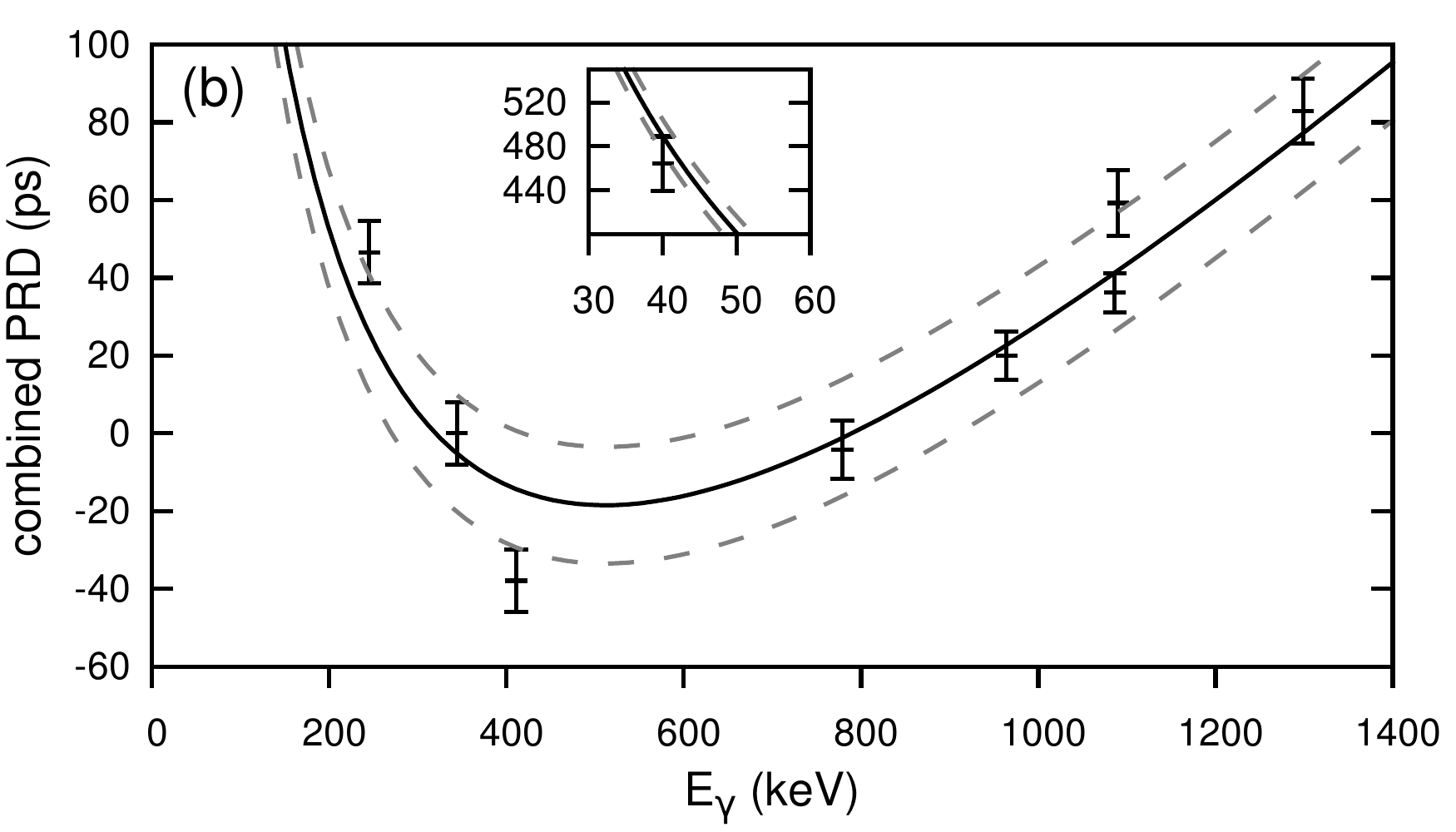}
\caption{PRD calibration points measured with a $^{152}$Eu source for
$^{172}$Hf, $^{174}$Hf (a), and $^{176}$Hf (b). The reference energy is
$E_\gamma = 344$ keV. The fitted calibration function is shown as a solid line.
The uncertainty of $\pm 10$ ps is indicated as shifted dashed lines. For better
comparison the data point at $E_\gamma = 40$ keV is shown as an inlay. See text
for more details.}\label{fig:PRDcalib}
\end{figure}

\begin{figure*}
\includegraphics[width=\linewidth]{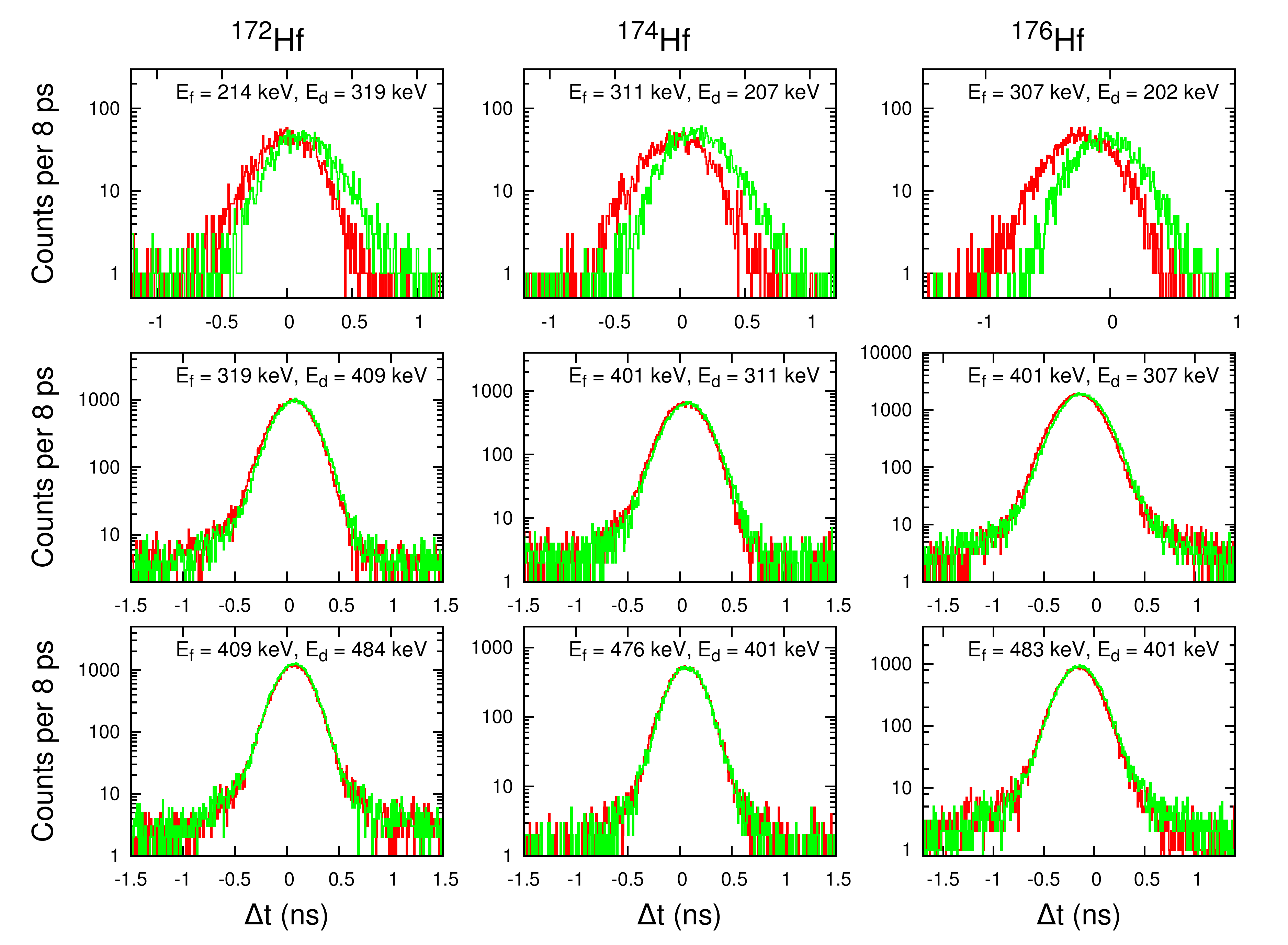}
\caption{(Color online) $\gamma$-$\gamma$ time difference spectra for direct 
decay-feeder
cascades in the nuclei $^{172}$Hf (left), $^{174}$Hf (center), and $^{176}$Hf
(right). The delayed spectra, with the decay gated on the stop branch, are 
shown in
gray (color: green). The antidelayed spectra, where the decay is gated on the 
start 
branch, are shown in black (color: red). The top row shows the time spectra for 
the $6^+_1 \rightarrow
4^+_1 \rightarrow 2^+_1$ cascade (with ce gate), the center row those for the
$8^+_1 \rightarrow 6^+_1 \rightarrow 4^+_1$ cascade, and the bottom those for
the $10^+_1 \rightarrow 8^+_1 \rightarrow 6^+_1$ cascade (both without ce
coincidence condition).}\label{fig:centroids}
\end{figure*}

For the measurement of the, previously unknown, shorter half-lives of
the 4$^+$, 6$^+$, and 8$^+$ states, TAC spectra of $\gamma$-$\gamma$
coincidences between two LaBr detectors were analyzed. LaBr coincidence
energy spectra are shown in Figure \ref{fig:Yspecs}. For these fast timing
measurements in the ps region the generalized centroid difference (GCD) method
was employed \cite{Reg13}, a refinement of the centroid shift method \cite{PhysRev.77.419}.
With the GCD method, the centroids of two time distributions are measured for a combination of transitions that mark 
the population (feeding transition) and the depopulation (decay transition) of a nuclear excited state. In 
the case of the present setup, if the half life of the excited state is shorter than 1ps, the decay can be considered 
prompt. If the life time is longer a shift of the centroid is observed. If the decay is gated on the stop (stop signal 
on the TAC), the result is a delayed time spectrum with a centroid $C_\mathrm{delayed}$ shifted to the right with 
respect to the prompt distribution. If the decay is gated on the start (start signal on the TAC), the result is an 
antidelayed time spectrum with centroid $C_\mathrm{antidelayed}$. A lifetime will show itself as a shift of 
$C_\mathrm{antidelayed}$ to the left with respect to the prompt position. The halflife of the excited state can be 
measured by measuring the difference between the delayed and antidelayed centroid if the centroid difference of two 
prompt signals for the respective energy combination of decay and feeder is known.
The necessary calibration of the prompt response
difference (PRD) was obtained from measurements with a $^{152}$Eu source, as
described in detail in \cite{Reg13}, using the calibration function given in 
the same
reference. Figure~\ref{fig:PRDcalib} shows the PRD calibration points and the
fitted function for the reference energy 344 keV for the two parts of
the experiment. An accuracy of \mbox{$\pm$ 10 ps} was adopted. Given
good statistics and peak-to-background ratio, this is the main contribution
to the experimental uncertainty.
The experimental observable
to be measured is the centroid difference
$\Delta C = C_\mathrm{delayed}-C_\mathrm{antidelayed}$
which is determined from the delayed and antidelayed
time spectra like the ones shown in Figure~\ref{fig:centroids}.
The difference between $\Delta C$ and the mean PRD, shifted to a given reference
energy, corresponds to twice the lifetime $\tau = \frac{T_{1/2}}{\ln(2)}$.
\begin{eqnarray}
2\tau &= \lvert \Delta C - \mathrm{PRD}(\Delta E_\gamma) \rvert
\label{eqn:2tau}\\
\mathrm{PRD}(\Delta E_\gamma) &= \mathrm{PRD}(E_\mathrm{feeder} -
E_\mathrm{decay} \label{eqn:PRDdE})
\end{eqnarray}

Analyzed time spectra which are measured using $\gamma$ detectors, like the
ones shown in Figure~\ref{fig:centroids}, necessarily contain background
from Compton scattering and crosstalk events. These are correlated in time
with the transitions gated on, but not in the same way as the full energy
peaks. This timing background is delayed with respect to the prompt response
for full energy events which leads to a systematic error in $\Delta C$.
Therefore a time correction related to the Compton background
contribution was performed in all cases, according to the procedure outlined in
reference \cite{Reg14}. The centroid of the background distribution at the
gate position was determined by interpolation. By assuming that the measured
distribution is the sum of the background distribution and the desired full
energy response, the centroid of the latter can be determined with the
additional knowledge of the peak-to-background ratio (ptb). The correction of course contributes
to the measurement uncertainty $\sigma_{T_{1/2}}$. That is

\begin{equation}
\sigma_{T_{1/2}} = \sqrt{ \sigma_{\Delta C}^2+ 
\sigma_\mathrm{PRD}^2+\left[\frac{\sigma(\Delta C - \Delta 
C_\mathrm{C.})}{\mathrm{ptb}}\right]^2}  \cdot 
\frac{\ln(2)}{2}\label{eqn:sigmaThalf}
\end{equation}

with $\sigma_{\Delta C}$ from the measurement of the centroid difference,
and $\sigma_{\mathrm{PRD}} = 10 \mathrm{ps}$ from the PRD calibration, and the 
uncertainty from the background correction, which includes the interpolated 
centroid difference of the Compton background $\Delta C_\mathrm{C.}$.
The uncertainties are those for
the actually measured quantity $2\tau$. The result is then scaled to the
half-life value via the rightmost factor in Eq. \ref{eqn:sigmaThalf}.
Dependent on the ptb and the time shift between the
measured CD and the interpolated time response of the Compton background $\Delta 
C_\mathrm{C.}$, the correction can be important.
For an estimation of the magnitude
of the correction and its effect on the uncertainty of the 
final half life we
give an example in numbers: In the case of the 311-401 coincidence in $^{174}$Hf
$\mathrm{ptb-ratio} \approx 5$. The absolute value of
the correction of $\Delta C$ is 2 ps, i.e. rather small. The
uncertainty of $\mathrm{T}_{1/2}$ is $\sigma_{\mathrm{T}_{1/2}} = 3.9 
\mathrm{ps}$ without
background correction and $\sigma_{\mathrm{T}_{1/2,\mathrm{bg}}}
= 4.8 \mathrm{ps}$ with background correction. 
In the case of the 311-417 keV coincidence in $^{174}$Hf, with 
$\mathrm{ptb-ratio} \approx 2$, the absolute value of the correction of $\Delta
C$ is about 16 ps. The uncertainties are $\sigma_{\mathrm{T}_{1/2}} = 
4.0
\mathrm{ps}$ and $\sigma_{\mathrm{T}_{1/2,\mathrm{bg}}} = 5.9 \mathrm{ps}$. 
From this example
it can be seen, that the background correction is very important, especially in
the second case. The effect on the uncertainty of the final half life, however,
is small compared to the contribution of the
overall PRD uncertainty of 10 ps.


It was possible to set a coincidence condition on the detection of a 2$^+_1$
$\rightarrow$ 0$^+_1$ electron for the $\gamma$-$\gamma$ timing analysis to
clean the LaBr spectra. This improved the peak-to-background ratio in the LaBr
energy gates used for the GCD analysis by a factor of up to 1.4 in the energy
region below 300 keV. An example is shown in Figure~\ref{fig:Y2plus}.
The statistics are of course drastically reduced. In the case of the strong
$4^+ \rightarrow 2^+$-transition, however, the systematic uncertainty due to
the background is the most severe among all transitions. Therefore this
approach was only used for the determination of the $4^+_1$ half-life.
Throughout the analysis only time spectra from coincidences between shielded
LaBr detectors were used, except in the determination of the lower half-life
limit of the $8^+_1$ states. Here the transition energies are greater than $400$
keV, i.e. above the region where crosstalk events play a significant role 
\cite{phdjm}.
The measurement of the centroid of a single distribution imposed an
uncertainty between $\pm$ 0.5 ps and $\pm$ 4 ps, depending on
available statistics. Time spectra of direct feeder-decay cascades for the
$4^+_1$, $6^+_1$, and $8^+_1$ states are shown in Figure~\ref{fig:centroids}.
The spectra were re-binned to \mbox{8 ps} per channel for the purpose of
display. The original resolution was 2 ps per channel. All measured half-lives
are summarized in the Tables \ref{tab:results_172Hf} -
\ref{tab:results_176Hf}. 

One merit of the GCD, which it inherits from the centroid shift
method \cite{PhysRev.77.419, Mach89}, is, that lifetimes $\tau_1, \tau_2,
\dots$ of intermediate levels connected by a cascade of several transitions
lead to a centroid shift about an effective lifetime $\tau'$ which corresponds
to the sum of the single lifetimes \cite{Mach89}
\[\tau' = \tau_1 + \tau_2 + \cdots .\]
This way it is possible to check the measured half-lives, and not least the
PRD calibration, for consistency by analyzing time spectra with
several coincidence conditions, i.e decay-feeder combinations. Where
possible this approach was also used (see Tab. \ref{tab:results_172Hf} -
\ref{tab:results_176Hf}).

\section{Experimental results\label{sec:results}}

All measured half-lives are shown in
Tables \ref{tab:BE2s} along with B(E2)
values. Conversion coefficients for the determination of B(E2) values
were calculated using BrIccFO \cite{Kibedi}, see Tab. \ref{tab:ICCs}.

\begin{figure*}
\includegraphics[width=\linewidth]{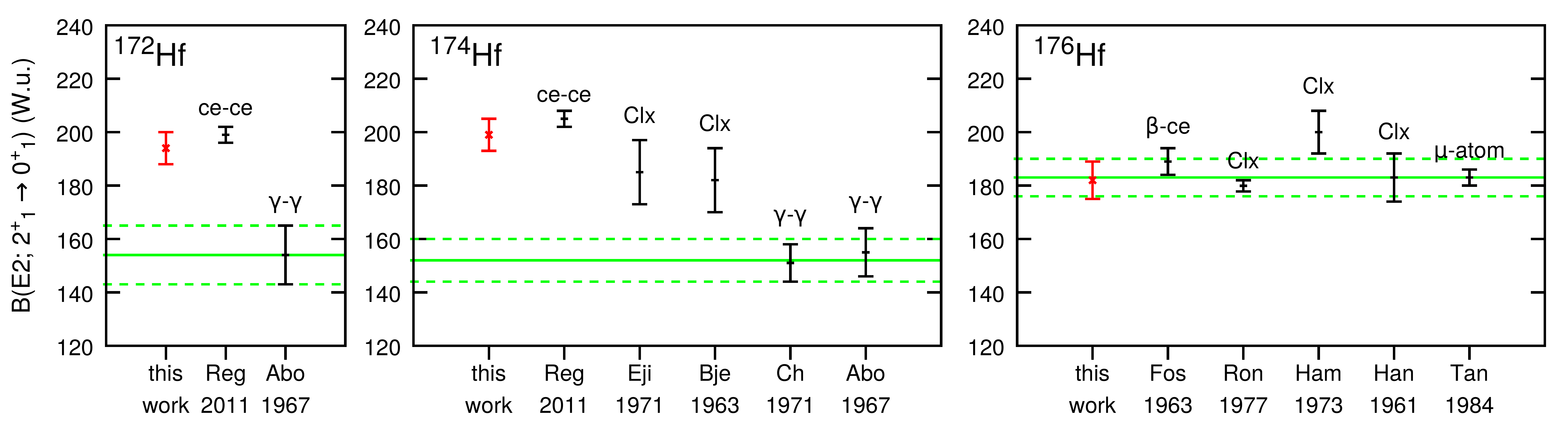}
\caption{(Color online) Comparison of measured $B(E2, 2^+_1 \rightarrow
0^+_1)$ values of $^{172}$Hf, $^{174}$Hf, and $^{176}$Hf. The respective adopted 
values found in the nuclear data
sheets \cite{NNDC172, NNDC174, NNDC176} are shown as a solid line with dashed
lines indicating the uncertainty. Other data are from
Han1961 \cite{Han61}, 
Bje1963 \cite{Bje63},
Fos 1963 \cite{Fos63},
Abo1967 \cite{Abo67},
Cha1971 \cite{Cha71},
Eji1971 \cite{Eji71},
Ron1977 \cite{Ron77},
Ham1973 \cite{Ham73},
Tan1984 \cite{Tan84},
and Reg2011 \cite{phdjm}. The experimental
method is indicated by ce-ce, $\gamma$-$\gamma$, $\beta$-ce (delayed
ce-ce, $\gamma$-$\gamma$ and $\beta$-ce coincidence fast timing), Clx
(Coulomb excitation), and $\mu$-atom (muonic atom
spectroscopy).}\label{fig:compare_BE2}                                          
           
\end{figure*}

\begin{table}
\caption{Measured half-lives in $^{172}$Hf. If there is more than one
intermediate level, and where the final result is an upper limit, the
effective half-life is given in the column T'$_{1/2}$. The adopted half-life
values are set in bold face. The values determined with the GCD method
are corrected for background contributions. For more details see
text.}
\begin{ruledtabular}
\begin{tabular}{cccrlrlc}\label{tab:results_172Hf}
	&E$_\mathrm{decay}$	&E$_\mathrm{feeder}$	&T'$_{1/2}$
&\bm{$\mathrm{T}_{1/2}$	}	\\
state	&(keV)			&(keV)			&(ps)		&(ps)	
		\\[1ex]\hline
2$^+_1$	&95			&214			&	
&\textbf{1250(40)}		\\[1ex]
4$^+_1$	&214			&319			&	
&\textbf{66}\textbf{(5)}	\\
	&214			&409			&78(6)		&63(12)	
		\\
	&214			&484			&88(13)		&70(17)	
		\\
	&214			&543			&95(18)		&73(20)	
		\\[1ex]
6$^+_1$	&319			&409			&	
&\textbf{15}\textbf{(8)}	\\      
 	&319			&484			&18(9)		&15(12)	
		\\[1ex]
8$^+_1$	&409			&484 			&3(7)&\bm{$<10$}\\

\end{tabular}
\end{ruledtabular}
\end{table}

\begin{table}
\caption{Same as in Table \ref{tab:results_172Hf} but for $^{174}$Hf.
}
\begin{ruledtabular}
\begin{tabular}{cccrlrlc}\label{tab:results_174Hf}
	&E$_\mathrm{decay}$	&E$_\mathrm{feeder}$	&T'$_{1/2}$
&\bm{$\mathrm{T}_{1/2}$	}	\\
state	&(keV)			&(keV)			&(ps)		&(ps)	
		\\[1ex]\hline
2$^+_1$	&91			&			&	
&\textbf{1280(40)}		\\[1ex]
4$^+_1$	&207			&311			&	
&\textbf{77}\textbf{(5)}	\\
	&207			&401			&101(6)		&85(9)	
		\\
	&207			&476			&99(15)		&84(15)	
		\\
	&207			&941			&65(47)		&49(52)	
		\\
	&207			&1252			&		&72(47)	
		\\[1ex]
6$^+_1$	&311			&401			&	
&\textbf{16}\textbf{(5)}	\\      
 	&311			&476			&22(6)		&18(8)	
		\\[1ex]
8$^+_1$	&401			&476			&5(5)	
&\bm{$<10$}			\\
\end{tabular}
\end{ruledtabular}
\end{table}

\begin{table}
\caption{Same as in Table \ref{tab:results_172Hf} but for $^{176}$Hf.
}
\begin{ruledtabular}
\begin{tabular}{cccrlrlc}\label{tab:results_176Hf}
	&E$_\mathrm{decay}$	&E$_\mathrm{feeder}$	&T'$_{1/2}$
&\bm{$\mathrm{T}_{1/2}$}		\\
state	&(keV)			&(keV)			&(ps)		&(ps)	
			\\[1ex]\hline
2$^+_1$	&88			&202			&
&\textbf{1470(60)}		\\[1ex]
4$^+_1$	&202			&307			&	
&\textbf{90}\textbf{(6)}		\\
	&202			&401			&110(6)		&93(9)	
			\\
	&202			&483			&114(6)		&92(12)	
			\\
	&202			&736			&109(8)		&93(11)	
			\\
	&202			&1043			&		&97(13)	
			\\[1ex]
6$^+_1$	&307			&401			&	
&\textbf{17(6)}					\\      
	&307			&483			&23(7)		&18(11)	
			\\[1ex]
8$^+_1$	&401			&483			&6(9)	
&\bm{$<15$}				\\
\end{tabular}
\end{ruledtabular}
\end{table}

\begin{table*}
\caption{Adopted values of the measured half-lives (see
Tables \ref{tab:results_172Hf}-\ref{tab:results_176Hf}) and the
corresponding B(E2) value for the transition to the next lower lying state in
the ground state rotational band. 
}
\begin{ruledtabular}
\begin{tabular}{ccccccc}\label{tab:BE2s}
	&\multicolumn{2}{c}{$^{172}$Hf}			&\multicolumn{2}{c}{$^{174}$Hf}		&\multicolumn{2}{c}{$^{176}$Hf} \\
	&T$_{1/2}$		&B(E2)			&T$_{1/2}$	& B(E2)			&T$_{1/2}$		&B(E2)		\\
state	&(ps)			&(W.u.)			&(ps)		&(W.u.)			&(ps)			&(W.u.)		\\[1ex]\hline
2$^+_1$	&1250(40)		&194(6)			&1280(40)	&199(6)	
	&1470(60)		&182(7)		\\[1ex]
4$^+_1$	&66(5)			&274(18)		&77(5)		&270(20)		&90(6)			&251(18)	\\[1ex]
6$^+_1$	&15(8)			&$177(^{+81}_{-42})$	&16(5)		&$197(^{+90}_{-47})$	&17(6)			
&$195(^{+106}_{-51})$	\\[1ex]
8$^+_1$	&{\bm{$<$}10}		&{\bm{$>$}85}		&{\bm{$<$}10}	&{\bm{$>$}91}		&{\bm{$<$}15}		&{\bm{$>$}60}	\\
\end{tabular}
\end{ruledtabular}
\end{table*}

Figure~\ref{fig:compare_BE2} shows a comparison of the results from this work
with results from earlier measurements of $B(E2, 2^+_1 \rightarrow 0^+_1)$ in
the three hafnium isotopes which were investigated. For $^{172}$Hf there is
one other measurement which also applied the $\gamma\gamma$ fast timing method,
by \citeauthor{Abo67} \cite{Abo67} who gives a value of $T_{1/2} = 1.55(10)$
ns. This results in $B(E2, 2^+_1 \rightarrow 0^+_1) = 154(11)$ W.u., which is
several $\sigma$s lower than the value from this work. Another measurement,
performed using ce-ce fast timing with the Cologne Double Orange Spectrometer,
resulted in $B(E2, 2^+_1 \rightarrow 0^+_1) = 199(3)$ W.u. \cite{phdjm}, in
agreement with the value found in this work.
The situation is similar for $^{174}$Hf, for which the Nuclear Data Base gives
an adopted value of $B(E2, 2^+_1 \rightarrow 0^+_1) = 152(8)$ W.u.
\cite{NNDC174}. Two sources are cited which also used $\gamma$-$\gamma$ fast
timing. From those the adopted value is calculated. One measurement, however,
used Coulomb excitation \cite{Eji71}, yielding a significantly larger
value of $B(E2, 2^+_1 \rightarrow 0^+_1) = 184(13)$ W.u. This is in agreement
with the value determined in this work. A ce-ce fast timing measurement resulted
in $B(E2, 2^+_1 \rightarrow 0^+_1) = 201(3)$ W.u., also much higher than the
value adopted by the Nuclear Data Sheets. For $^{176}$Hf the value of this
work agrees well with the one adopted by the NNDC. In this case the sources
do not contain any $\gamma$-$\gamma$ fast timing experiments from before 1972. 
The
measurements which resulted in the low values cited above were all performed in
the late 1960s, and used NaI detectors. The energy resolution of these detectors
is much worse than that of LaBr detectors used today. Under such circumstances
it is difficult to control and estimate the background contributions,
especially for transitions as low in energy as the $2^+_1 \rightarrow 0^+_1$
transition in deformed nuclei, which typically lie around or below 100 keV.
For this reason we will adopt the $B(E2, 2^+_1 \rightarrow 0^+_1)$ values
measured in this work for our further discussion. The other half-lives and
absolute transition strengths measured in this work are measured for the first
time. 

\begin{figure}
\begin{center}
\begin{tabular}{ccc}
 \includegraphics[width=1\linewidth]{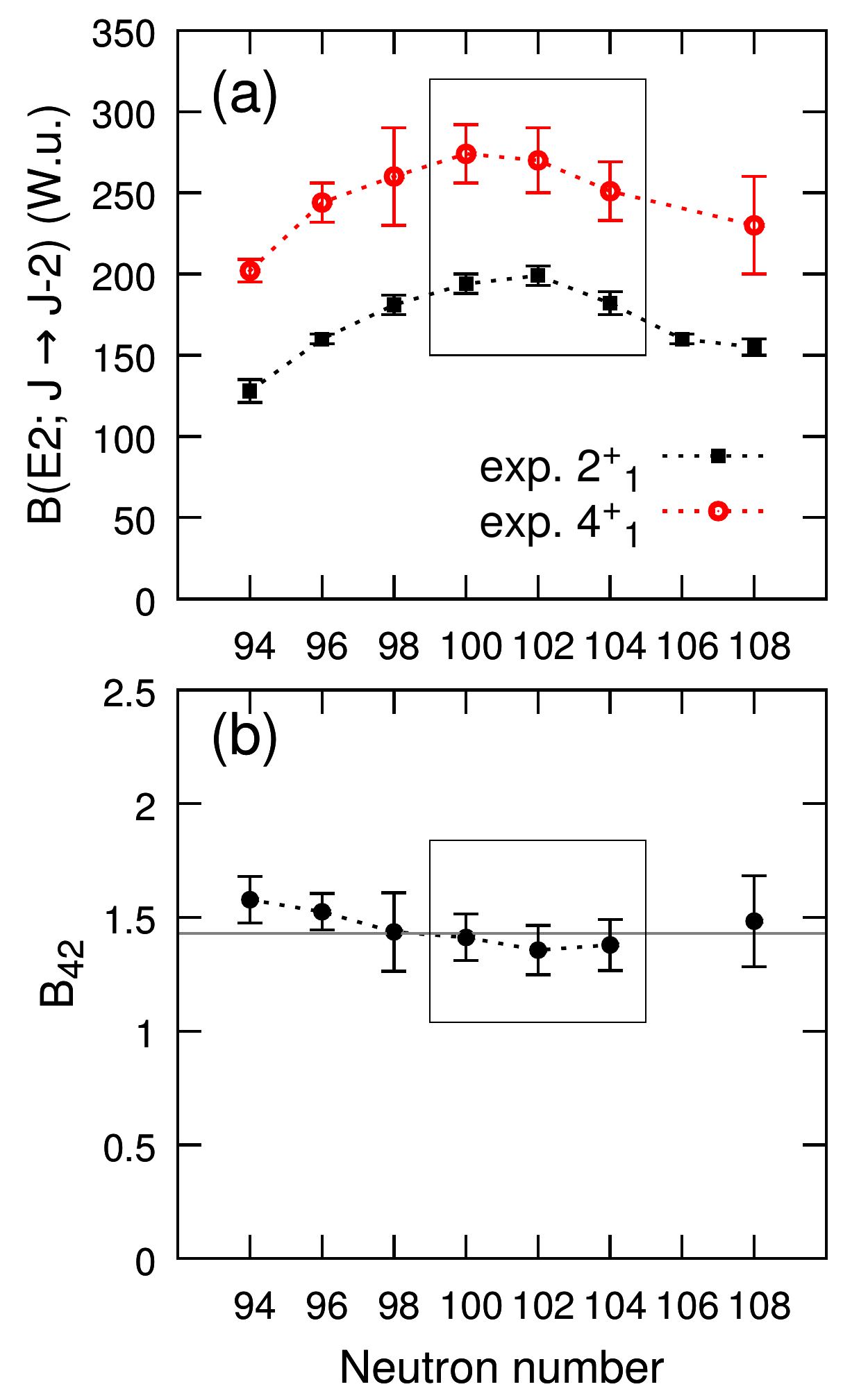} &
\end{tabular}
\end{center}
\caption{(Color online) Experimental $B(E2; J \rightarrow J-2)$ values in
hafnium isotopes (Z=72) (a) and the ratio $B_{42} = B(E2, 4^+_1 \rightarrow 2^+_1)/B(E2, 
2^+_1 \rightarrow 0^+_1)$ (b). Values measured in this work are marked with a box.
Other values are taken from the Nuclear Data Sheets \cite{NNDC166, NNDC168,
NNDC170, NNDC178, NNDC180}.}
\label{fig:exp_only_BE2}
\end{figure}

\begin{table}
\caption{Internal conversion coefficients (ICCs) from \mbox{BrIccFO} \cite{Kibedi} used for the calculation of B(E2) 
values.}
\begin{ruledtabular}
\begin{tabular}{cccc}\label{tab:ICCs}
				&\multicolumn{3}{c}{internal conversion coefficient $\alpha$}\\
transition 			&$^{172}$Hf		&$^{174}$Hf		&$^{176}$Hf	\\[1ex]\hline
$2^+_1 \rightarrow 0^+_1$	&4.32(6)		&5.12(8)		&5.86(9)		\\
$4^+_1 \rightarrow 2^+_1$	&0.230(4)		&0.256(4)		&0.278(4)		\\
$6^+_1 \rightarrow 4^+_1$	&0.0660(10)		&0.0711(10)		&0.0739(11)		\\
$8^+_1 \rightarrow 6^+_1$	&0.0327(5)		&0.0345(5)		&0.0345(5)		

\end{tabular}
\end{ruledtabular}
\end{table}
The new $B(E2)$ values complete the picture of evolution of quadrupole 
transition strength in the ground state band of mid-shell hafnium isotopes. The
result is a smooth increase from low neutron numbers $N$ towards mid-shell and a
subsequent smooth decrease (see Fig. \ref{fig:exp_only_BE2} (a)).
It is well established that the investigated hafnium isotopes display the 
characteristics of an axially deformed rotor.
In such nuclei the ratio \mbox{$B_{42} = B(E2, 4^+_1 \rightarrow 2^+_1)/B(E2, 
2^+_1 \rightarrow 0^+_1)$} can be calculated as the ratio of 
Clebsch-Gordan coefficients (Alaga rules), which yields  $B_{42} = 1.43$. As can
be seen in Figure~\ref{fig:exp_only_BE2} (b), the newly measured transition
strengths fit well into the systematics for rigid rotors, as expected.

\section{Calculation and discussion\label{sec:calculations}}

\subsection{Description of the model and discussion of mean-field results}

To help understand the data, we have performed theoretical calculations of
spectroscopic properties by
employing the recently proposed methodology of Refs.~\cite{Nom08,Nom11}. 
The essential idea of the method is to determine the Hamiltonian of an
appropriate version of the IBM by computing the bosonic deformation
energy surface so that it reproduces, in a way described below,
the basic topology of the  deformation energy
surface of the many-fermion system computed microscopically with the HFB method. 
The resultant Hamiltonian is used to calculate energy levels and wave
functions of excited states. 

\begin{figure*}
\begin{center}
\begin{tabular}{ccc}
 \includegraphics[width=0.33\linewidth]{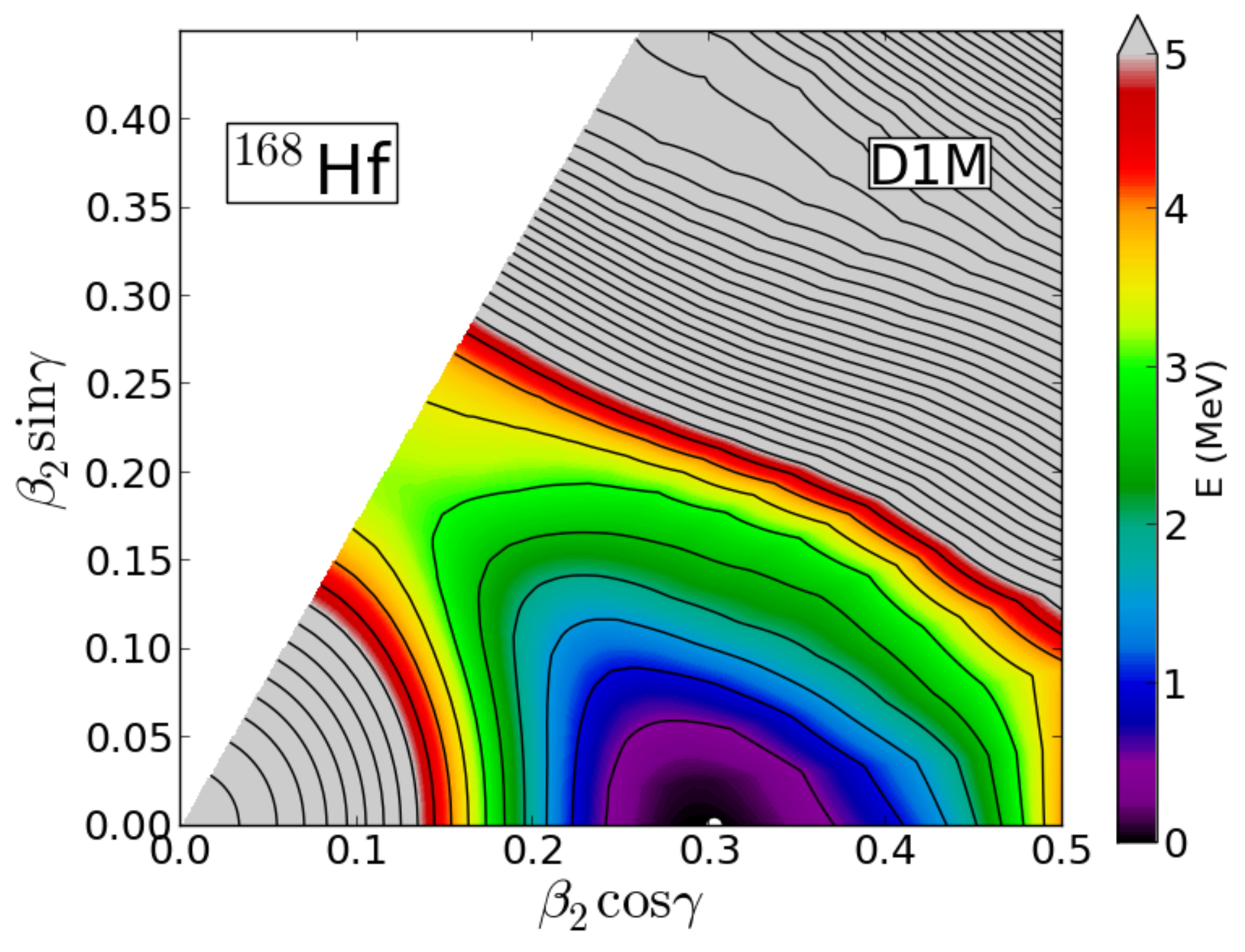} &
 \includegraphics[width=0.33\linewidth]{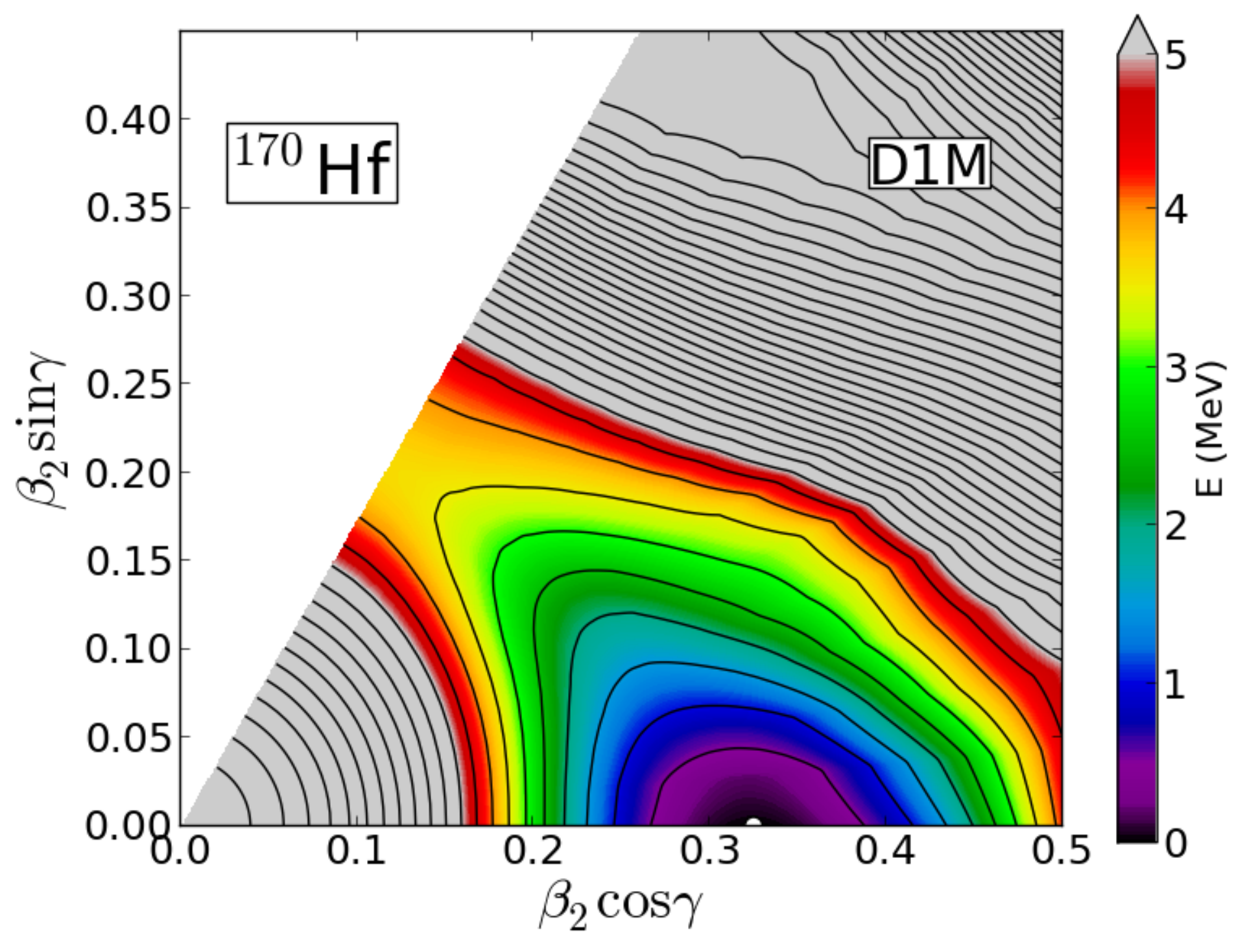} &
 \includegraphics[width=0.33\linewidth]{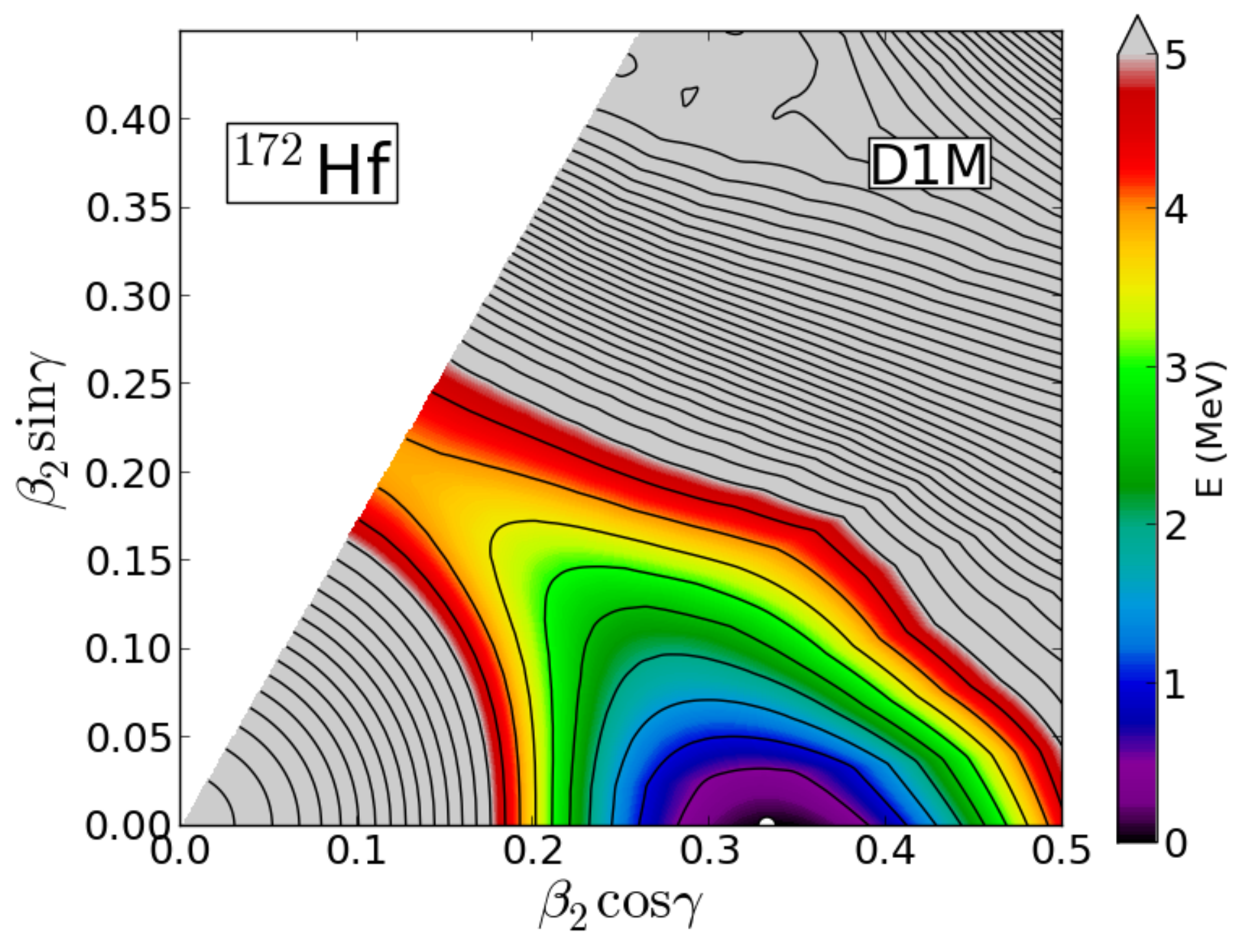} \\
 \includegraphics[width=0.33\linewidth]{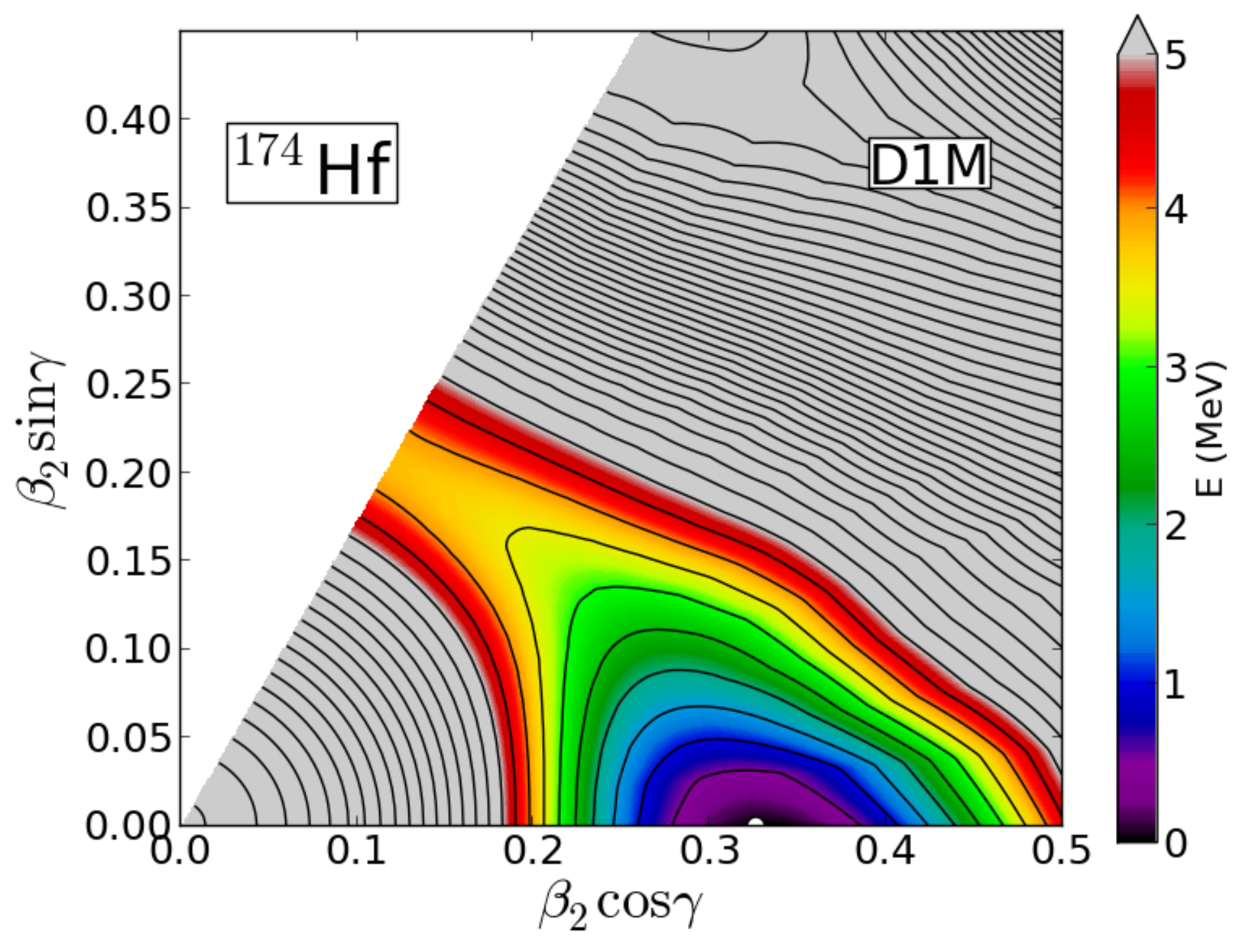} &
 \includegraphics[width=0.33\linewidth]{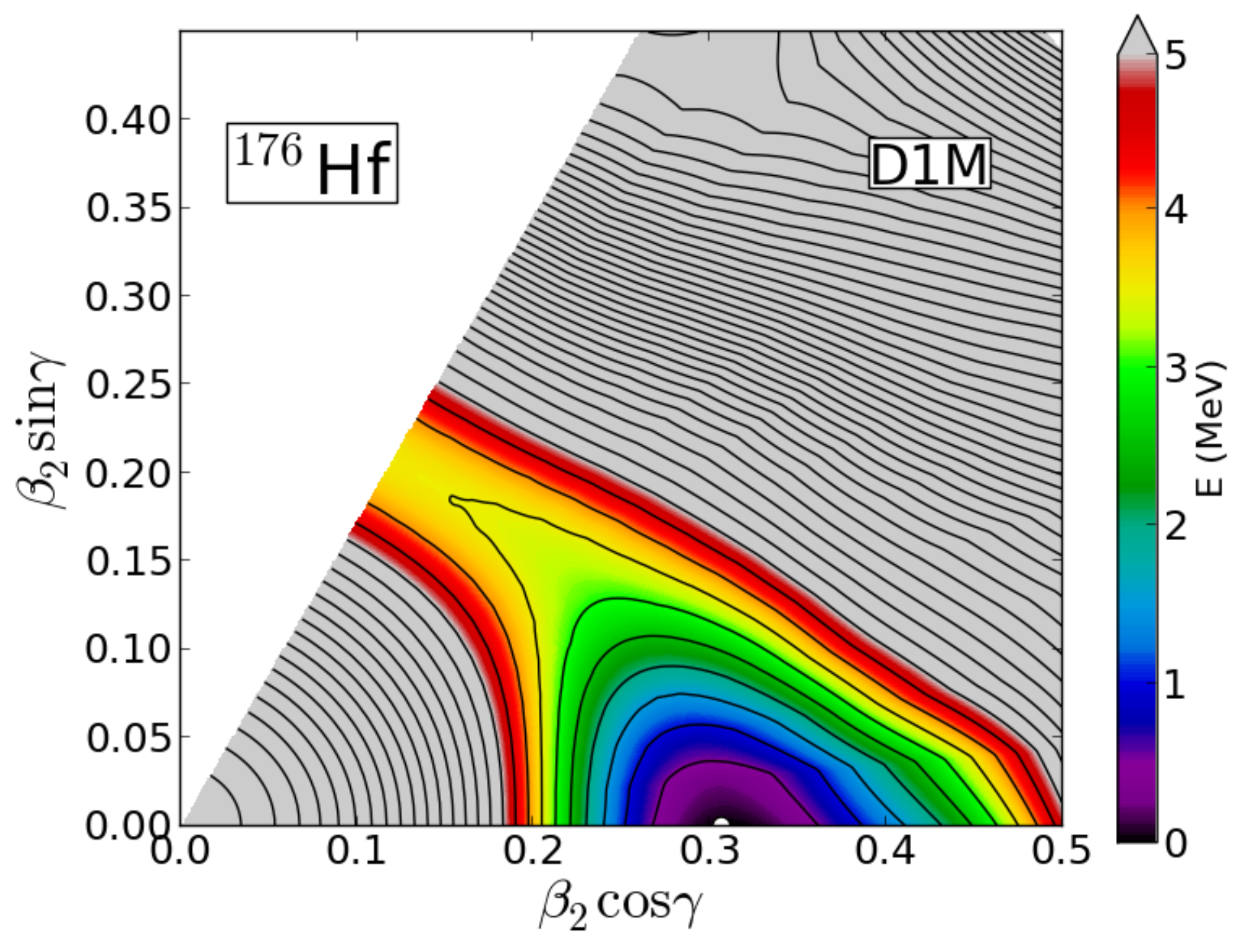} &
 \includegraphics[width=0.33\linewidth]{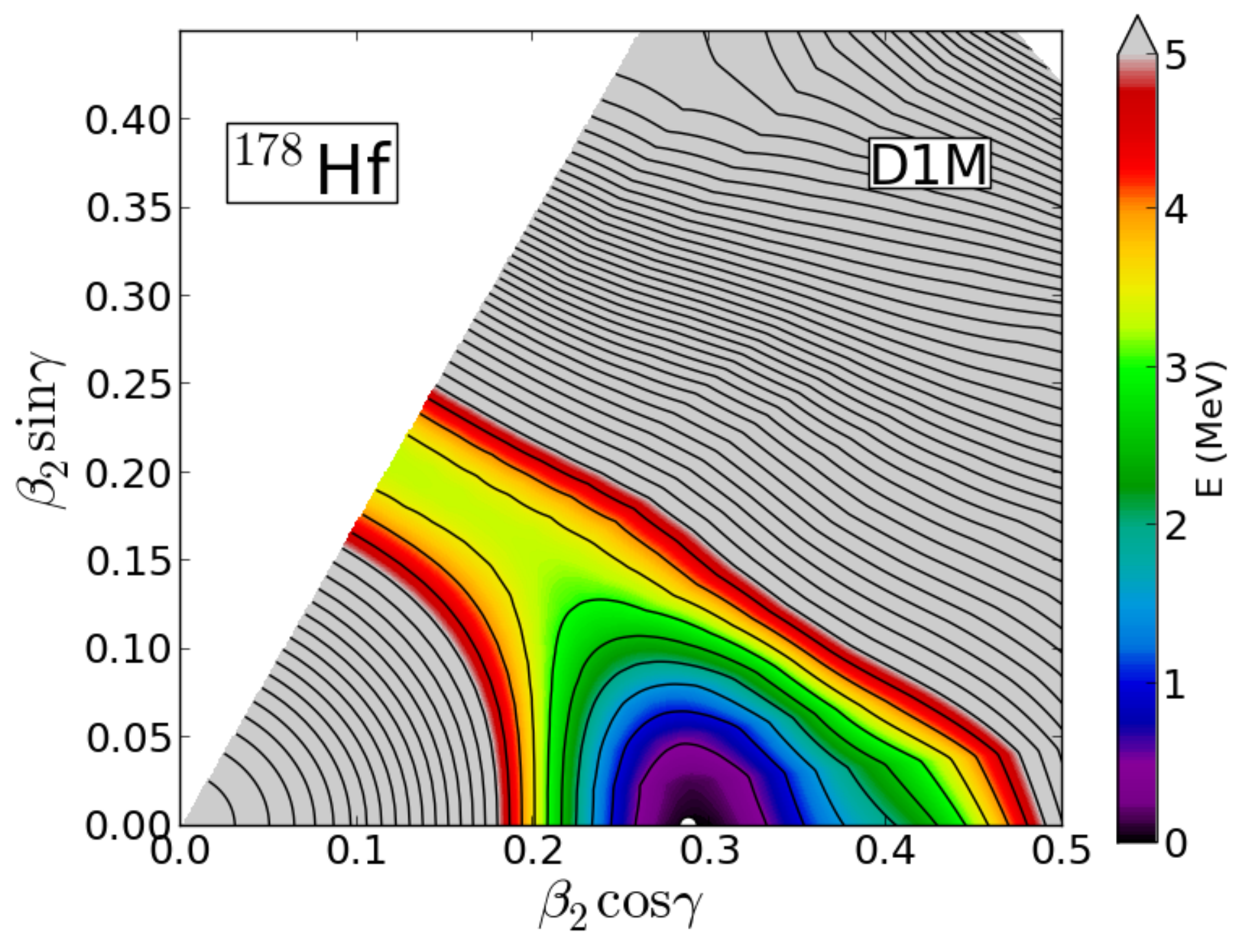} 
\end{tabular}
\end{center}
\caption{(Color online) Contour plots of the deformation energy surfaces in terms of the
quadrupole deformations $\beta_2$ and $\gamma$ 
 for the nuclei $^{168-178}$Hf, obtained from the Gogny HFB
 calculations using the D1M interaction. The color scale varies in
 steps of 0.25 MeV and the contour lines are drawn in steps of 0.5 MeV. The range of the plot is
 $0.0\leqslant\beta\leqslant 0.5$ and $0^{\circ}\leqslant\gamma\leqslant
 60^{\circ}$. The absolute minimum is identified 
 by an open circle. }
\label{fig:pes}
\end{figure*}

Our starting point is the microscopic calculation of the deformation energy
surface within the self-consistent mean-field model. 
We have performed, for each individual Hf
nucleus, a set of the constrained Hartree-Fock-Bogoliubov (HFB)
calculations, and obtained the deformation energy surfaces in terms of
the quadrupole collective coordinates $\beta$ and $\gamma$ \cite{BM}. 
It is also possible to parametrize the energy surfaces using the
quadrupole moments $Q_{20}$ and $Q_{22}$ related to
the $\beta$ and the $\gamma$ variables by 
$\beta=\sqrt{\frac{4\pi}{5}}\frac{Q}{A\langle
r^2\rangle}$ and $\gamma=\tan^{-1}\frac{Q_{22}}{Q_{20}}$ with
$Q=\sqrt{Q_{20}^2+Q_{22}^2}$ (see Ref.~\cite{Nom11} for more details).
In the definition of $\beta$ we use the mean-squared radius $\langle
r^2\rangle$ evaluated with the corresponding HFB state. 
Throughout this work, the D1M parametrization 
of the Gogny energy density functional \cite{D1M} is employed for the
effective nucleon-nucleon interaction.

In Fig.~\ref{fig:pes}, the deformation energy surfaces obtained from
the Gogny-D1M HFB calculations for the 
$^{168-180}$Hf nuclei, are plotted in terms of the $\beta$ and $\gamma$
deformations. 
We limit the plot to the range $0.0\leqslant\beta\leqslant 0.5$ and $0^{\circ}\leqslant\gamma\leqslant
 60^{\circ}$, because it is the relevant scope for our purposes. 
Energy surfaces for $^{166}$Hf and $^{180}$Hf are not shown because
they are quite similar in topology to the ones of the adjacent nuclei $^{168}$Hf and
$^{178}$Hf, respectively. 
In general, the energy minimum is located around $\beta=0.3$ on the
$\gamma=0$ axis, being characteristic of an axially deformed prolate rotor. 
While any significant change in the topology is visible from the
microscopic energy surface, the minimum appears to be steeper in
both $\beta$ and $\gamma$ directions for heavier Hf isotopes. 
Although not shown here, the results with D1S look quite similar to the D1M ones.
However, as compared to the D1M results, the energy minima of the D1S energy
surfaces are steeper both in $\beta$ and $\gamma$ directions than in the D1M case. 
In Table \ref{tab:pes} we observe that for the
nuclei $^{170-178}$Hf  the HFB energy of the minimum (relative to the energy of the spherical 
configuration ($\beta,\gamma$)=(0,0) and denoted as $E_{min}$)  is generally
around 1.2-1.7 MeV smaller in magnitude in the D1M case than the D1S one.

\begin{table}
\caption{
The HFB energy at the minimum (relative to the spherical
 configuration, and denoted by $E_{min}$) as well as  the position of the minimum $\beta_{2,min}$ are 
 given for both parametrizations of the Gogny force and the $^{170-178}$Hf nuclei.
}
\begin{ruledtabular}
\begin{tabular}{ccccc}\label{tab:pes}
	&\multicolumn{2}{c}{$E_{min}$ (MeV)} &\multicolumn{2}{c}{$\beta_{2,min}$}		\\
	&D1S		&D1M			&D1S	&
 D1M			\\
\hline
$^{170}$Hf	& -12.214 & -10.944 & 0.339 & 0.325 \\
$^{172}$Hf	& -13.788 & -12.277 & 0.360 & 0.332 \\
$^{174}$Hf	& -14.724 & -13.061 & 0.353 & 0.326 \\
$^{176}$Hf	& -15.068 & -13.385 & 0.320 & 0.307 \\
$^{178}$Hf	& -14.960 & -13.328 & 0.301 & 0.288 \\
\end{tabular}
\end{ruledtabular}
\end{table}

\begin{figure}
\begin{center}
 \includegraphics[width=\columnwidth]{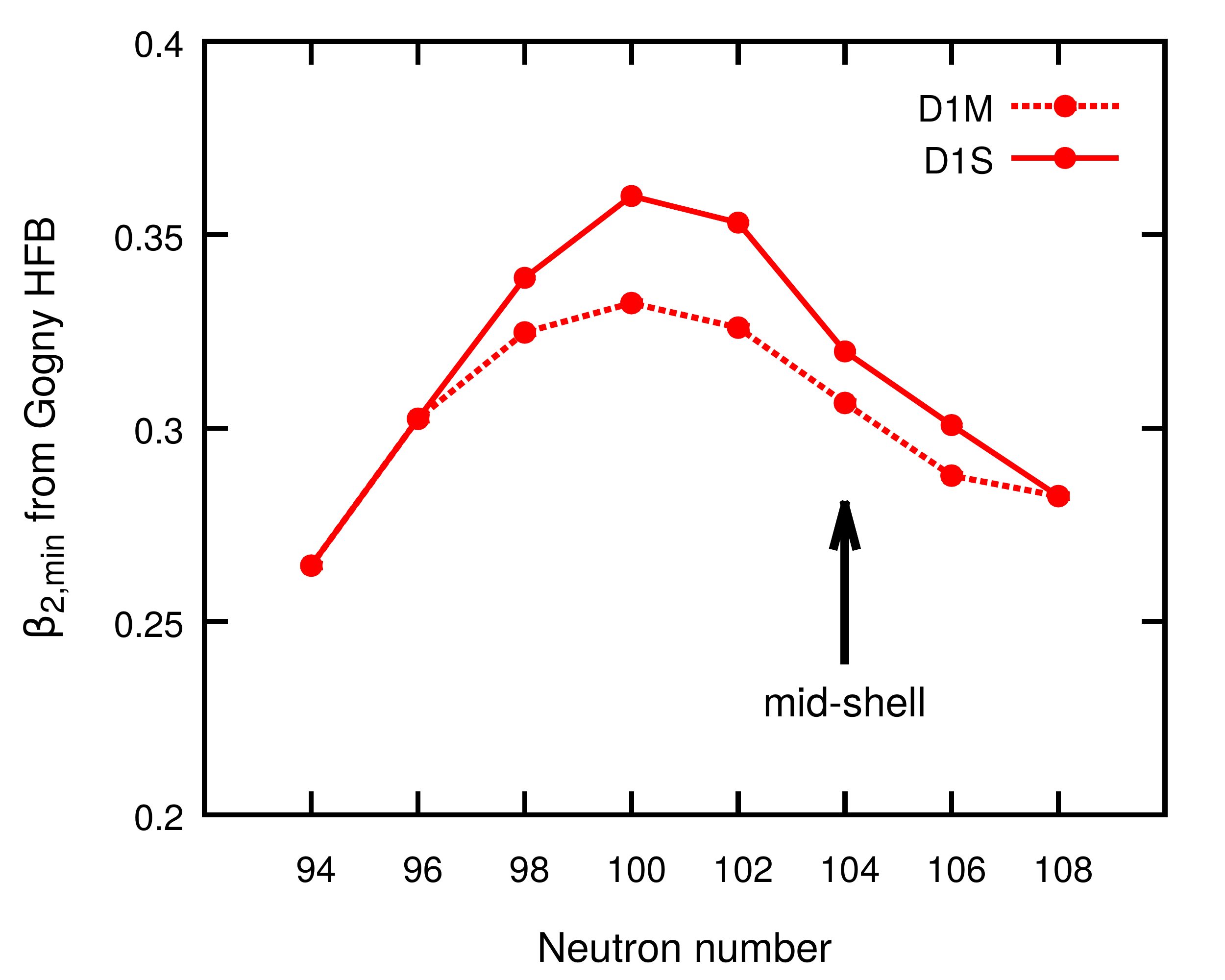}
\end{center}
\caption{(Color online) The $\beta_{2,min}$ values, where the HFB energy has 
its absolute minimum are plotted as a function of neutron number for the
$^{166-180}$Hf isotopes. Solid and 
 dotted curves represent the results from D1S and D1M interactions,
 respectively.}
\label{fig:theo_beta}
\end{figure}

In Fig.~\ref{fig:theo_beta} the $\beta$ value at  
the absolute minimum of the microscopic energy surface (denoted hereafter as 
$\beta_{2,min}$) is plotted as a function of 
neutron number. It exhibits a parabolic behavior with its maximum at $N=100$ instead
of the mid-shell value $N=104$. 
The D1S results are generally larger
than the D1M ones but in both cases they show their maximum value at $N=100$. 
However, as observed in  Table~\ref{tab:pes}, the minimum energy $E_{min}$
reaches its maximum  at  mid-shell for both the D1S and the D1M sets. 


For the boson calculation, we employ the following proton-neutron IBM (IBM-2) Hamiltonian: 
\begin{eqnarray}
\label{eq:bh}
 \hat H_{\textnormal{IBM}}=\epsilon(\hat n_{d\pi}+\hat n_{d\nu})+\kappa\hat Q_{\pi}\cdot\hat
  Q_{\nu}+\kappa^{\prime}\hat L\cdot\hat L, 
\end{eqnarray}
where the first and the second terms stand for the $d$-boson number
operator and the quadrupole-quadrupole interactions, respectively. They are given by the expressions
$\hat n_{d\tau}=d^{\dagger}_{\tau}\cdot\tilde d_{\tau}$ and
$\hat Q_{\tau}=s^{\dagger}_{\tau}\tilde 
d_{\tau}+d^{\dagger}_{\tau}s^{\dagger}_{\tau}+\chi_{\tau}[d^{\dagger}_{\tau}\times\tilde
d_{\tau}]^{(2)}$ with $\tau$ being either $\pi$ (proton) or $\nu$ (neutron). 
The third term is relevant for rotational bands and it is shown \cite{Nom11} to be necessary for the
description of deformed nuclei. It is given in terms of the total angular momentum operator
$\hat L=\hat L_\pi+\hat L_{\nu}$  with $\hat
L_{\tau}=\sqrt{10}[d^{\dagger}_{\tau}\times\tilde d_{\tau}]^{(1)}$.

The most general IBM-2 Hamiltonian contains many more terms
and parameters than the one in Eq.~(\ref{eq:bh}). 
The present IBM-2 Hamiltonian is rather simple compared with a general
Hamiltonian, but contains the minimal number of interaction
terms relevant for the description of low-lying quadrupole states. 
The Hamiltonian parameters are determined by mapping the HFB energy 
surface onto the IBM one. 
Some technical details of the procedure are described in
Appendix \ref{sec:mapping}. In
Table \ref{tab:IBMpara} we tabulate the  IBM-2 parameters determined by the mapping procedure for
the $^{166-180}$Hf isotopes. 

\begin{table}[htb!]
\caption{\label{tab:IBMpara}%
The parameters for the IBM Hamiltonian $\hat H_{\rm IBM}$ of
 Eq.~(\ref{eq:bh}), as well as the proportionality coefficient
 $C_{\beta}$ for the deformation parameter $\beta$, obtained from the mapping of the HFB to the IBM energy
 surfaces for $^{166-180}$Hf. 
}
\begin{ruledtabular}
\begin{tabular}{cccccccc}
\textrm{$A$}&
\textrm{$\epsilon$}&
\textrm{$-\kappa$}&
\textrm{$\chi_{\pi}$}&
\textrm{$\chi_{\nu}$}&
\textrm{$\kappa^{\prime}$}&
\textrm{$C_{\beta}$}&
\textrm{$e_B$}\\
\textrm{}&
\textrm{(keV)}&
\textrm{(keV)}&
\textrm{$\times 10^{3}$}&
\textrm{$\times 10^{3}$}&
\textrm{(keV)}&
\textrm{}&
\textrm{($e$fm$^2$)}\\
\colrule
166 & 474 & 281 & 213 & -813 & -9.95 & 3.65 & 15.3\\
168 & 439 & 285 & 213 & -729 & -11.9 & 3.20 & 16.3\\
170 & 451 & 292 & 241 & -895 & -13.0 & 2.96 & 15.9\\
172 & 333 & 281 & 454 & -878 & -10.5 & 2.96 & 15.5\\
174 & 238 & 267 & 303 & -768 & -6.46 & 3.00 & 14.3\\
176 & 166 & 242 & 237 & -723 & -2.00 & 3.30 & 12.9\\
178 & 97.7 & 258 & 709 & -1086 & -0.952 & 3.50 & 12.3\\
180 & 88.0 & 273 & 359 & -801 & -3.36 & 3.60 & 13.5\\
\end{tabular}
\end{ruledtabular}
\end{table}

Having all the relevant parameters at hand, the Hamiltonian is diagonalized to obtain
energies and wave functions of the excited states. 
For the numerical diagonalization of the IBM-2 Hamiltonian and the calculations
of the E2 transition rates, the computer
program NPBOS \cite{NPBOS} has been used. 
Using the resultant wave functions, the
transition probabilities between the states are calculated. 
In particular, the $B$(E2;$J\rightarrow J^{\prime}$) value is obtained
by 
\begin{eqnarray}
 B({\textnormal{E2}};J\rightarrow
  J^{\prime})=\frac{1}{2J+1}|\langle J^{\prime}||\hat
  T^{(E2)}||J\rangle|^2, 
\end{eqnarray}
where $J$ and $J^{\prime}$ are the total angular momenta of the initial
and the final states of the transition, respectively. 
The E2 operator is written as $\hat T^{(E2)}=e_{\pi}\hat
Q_{\pi}+e_{\nu}\hat Q_{\nu}$, where $\hat Q_{\tau}$ is the quadrupole
operator defined in Eq.~(\ref{eq:bh}) and the same values of
$\chi_{\pi,\nu}$ parameters are used. The parameter
$e_{\tau}$ is the boson effective charge for proton and neutron. 
Here we assume that the effective charge is the same for protons and
neutrons, $e_{\pi}=e_{\nu}\equiv e_{B}$. 
In most of the previous IBM calculations, 
a fixed value of the effective
charge $e_{B}$, that is determined by fitting
to the experimental data for the $B$(E2) values, is used for all members of the isotopic chain. 

\subsection{Energy levels}

\begin{figure}
\begin{center}
 \includegraphics[width=\columnwidth]{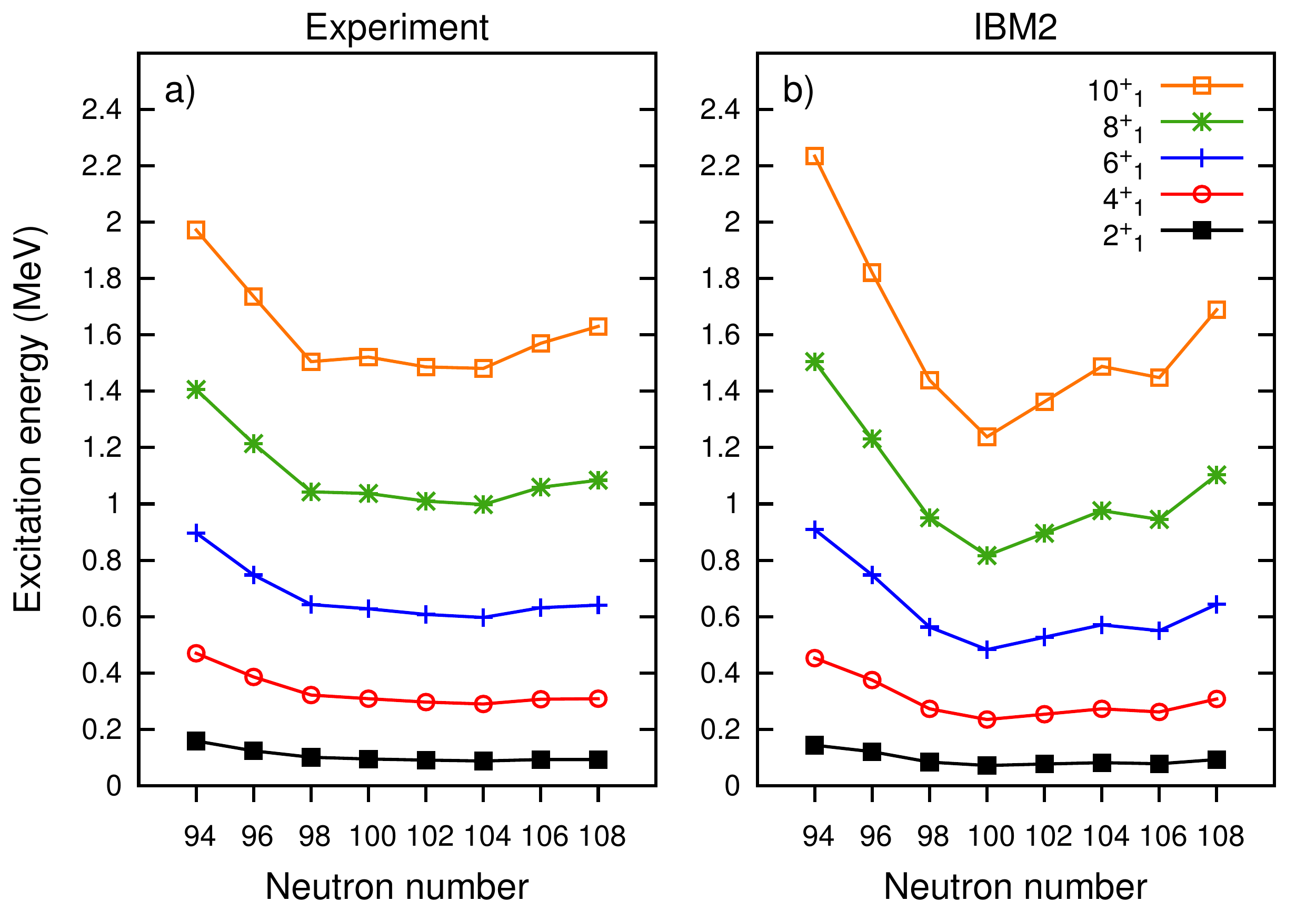}
\end{center}
\caption{(Color online) Experimental (a) and the theoretical (D1M 
parametrization) (b) level 
 energies of the low-lying yrast $2^+$, $4^+$, $6^+$, $8^+$ and $10^+$ states as functions of neutron number. } 
\label{fig:theo_energies}
\end{figure}

\begin{figure}
\begin{center}
 \includegraphics[width=\columnwidth]{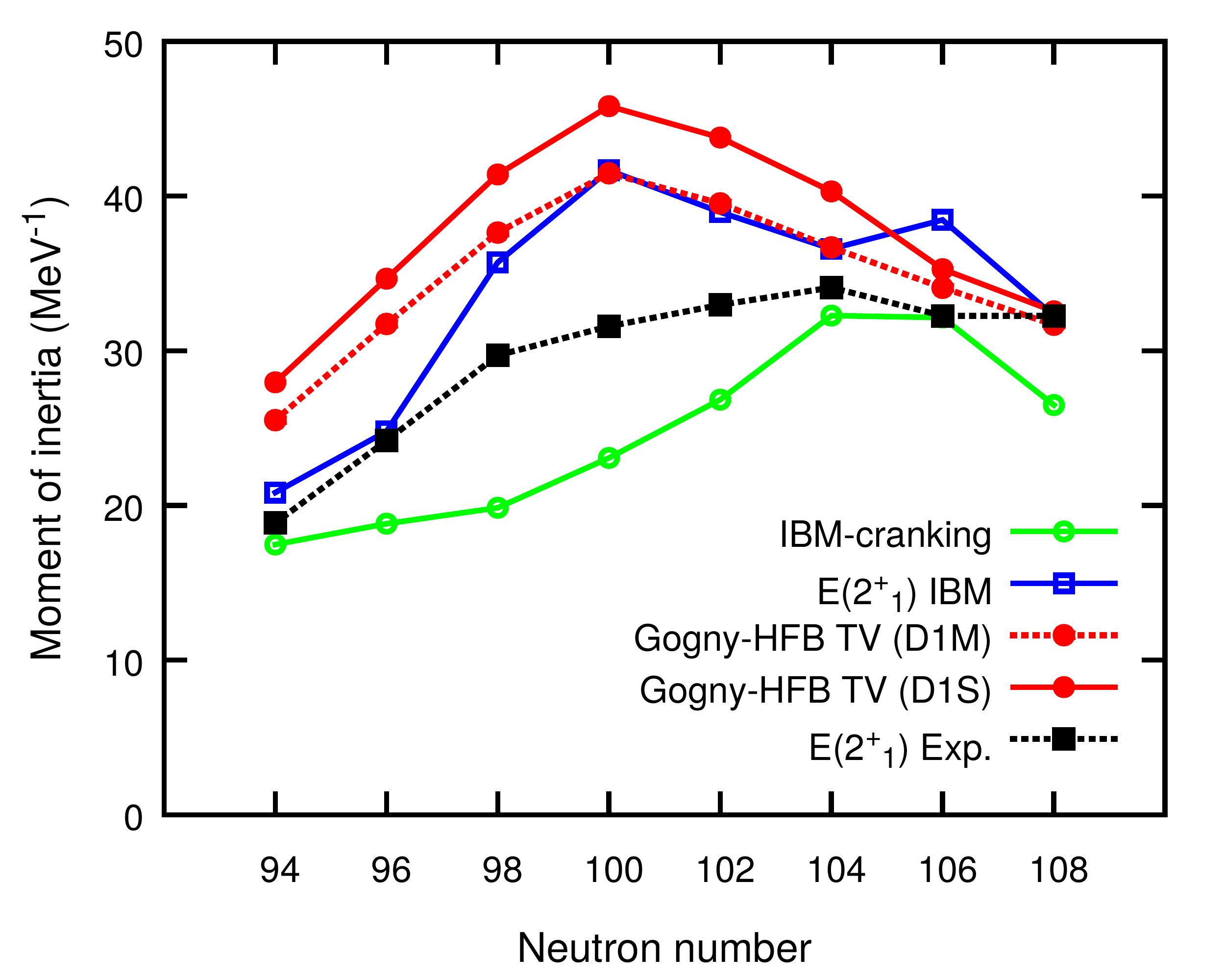}
\end{center}
\caption{(Color online) The moments of inertia (MOIs) of the ground state rotational band of
 the $^{166-180}$Hf nuclei obtained in different approaches are plotted
 as a function of neutron number. The MOIs are computed using the cranking calculations of
 the IBM in the coherent state ``IBM-cranking'' and  the self-consistent HFB cranking method that leads to the
 Thouless-Valatin MOI both with  the D1M ``Gogny-HFB TV (D1M)'' and
  the D1S  ``Gogny-HFB TV (D1S)'' parametrizations. Also shown are the MOI extracted from the $2^{+}_{1}$ 
 excitation energies of the IBM-2 ``$E(2^{+}_{1})$ IBM'' and the
 experiment ``$E(2^{+}_{1})$ Exp.'' using the rigid rotor formula. }
\label{fig:theo_moments}
\end{figure}

In order to confirm that the present framework gives a reasonable
description of the energy levels, we display in Fig.~\ref{fig:theo_energies} the experimental level
energies for the yrast $2^+$, $4^+$, $6^+$, $8^+$ and $10^+$ states and
the corresponding theoretical level energies calculated by the mapped IBM-2. 
Overall, our IBM-2 calculation follows the experimental trend satisfactorily, and 
reproduces the energies for each nucleus rather well at a quantitative
level. 
However, the energy levels have a minimum as a function of neutron number 
at mid-shell $N=104$ in the experiment, while the minimum is at $N=100$ in the 
theory. 

To illustrate this deviation of the $2^+_1$ excitation energy, 
we show in Fig.~\ref{fig:theo_moments} the moments of 
inertia (denoted as MOI) of the rotational band, obtained from the rotor formula
$E(J)\propto J(J+1)$ (with $J=0^+, 2^+, 4^+, \ldots$) and the $2^{+}$
energies of the IBM result (denoted as ``$E(2^{+}_1)$ IBM'' in the
figure) and the experiment (``$E(2^{+}_1)$ Exp.''). 
The moment of inertia, which is equal to $3/E(2^{+}_1)$, of the IBM is maximal at
$N=100$, while the experimental MOI at 
$N=104$ but changes much less than the theoretical value with neutron
number. 

The reason for the discrepancy in the systematics of the MOI between the IBM result and the experiment
can be due to the inclusion of the
$\hat L\cdot\hat L$ term in the IBM-2 Hamiltonian.  
To shed more light upon this point, we plot in the same figure the MOI obtained from the
cranking calculation in the Gogny HFB model using the Thouless-Valatin
(TV) formula \cite{TV} (denoted as ``Gogny-HFB TV'' in the figure for
the two different parametrizations D1S and D1M), and
the one calculated in the IBM without the $\hat L\cdot\hat L$ term
(denoted as ``IBM-cranking''). 
As described in Appendix \ref{sec:mapping}, 
the coefficient of the $\hat L\cdot\hat L$ term is determined so that
the ``IBM-cranking'' MOI becomes identical to the ``Gogny-HFB TV (D1M)'' one. 
For that reason, we do not plot the IBM cranking MOI with the $\hat
L\cdot\hat L$ term. 
In Fig.~\ref{fig:theo_moments}, one should see that the ``IBM-cranking''
MOI is peaked at $N=104$ or 106, consistent with the experimental MOI
while the ``Gogny-HFB TV (D1M)'' MOI is peaked at $N=100$ (the same applies for the D1S values). 
For the nuclei with $N\leq 102$, the difference between the ``IBM-cranking'' and the
``Gogny-HFB TV (D1M)'' MOIs is rather large as compared to the one for $N\geq
104$. 
Thus, the lowering of the energy due to the inclusion of the $\hat
L\cdot\hat L$ is much more significant in the $N\leq 102$ Hf nuclei than in
the $N\geq 104$ ones. 
Consequently, the maximum point of the ``$E(2^{+}_1)$ IBM'' MOI appears
at $N=100$ due to the inclusion of the $\hat L\cdot\hat L$ term.  
This correlates with the evolution  of the derived $\kappa^{\prime}$
parameter, presented in Table~\ref{tab:IBMpara}: for example, the parameter $\kappa^{\prime}$
for $^{172}$Hf is much larger in magnitude than that for $^{176,178}$Hf. 
We also note that the same conclusion would be extracted from the D1S
parametrization as it predicts the same systematics as the D1M set.

It is certainly worth noting that, contrary to the empirical trend, the
Thouless-Valatin MOI becomes maximal, irrespectively of the choice of
the parametrizations, not at the mid-shell $N=104$ and
that this systematics is well correlated with the
evolution of the $\beta_{2,min}$ in Fig.~\ref{fig:theo_beta}. 

On the other hand, it has been shown \cite{afanasjev00} that the moment
of inertia of the ground-state rotational band calculated in the
cranking model is rather sensitive to the details of the pairing
interaction. 
Therefore, it can be of interest that a small modification to the
relevant channel in the D1S and D1M functionals could lead to a
substantial improvement of the agreement with the experimental data. 

Finally, we also note that the calculation generally underestimates the
experimental level energies, in particular, for the nucleus $^{172}$Hf. 
The reason would be that the effect of including the $\hat L\cdot\hat L$
term is rather significant, as the TV MOI from the
Gogny-HFB calculation overestimates the experimental one considerably
(see Fig.~\ref{fig:theo_moments}).

\subsection{E2 transitions\label{sec:E2}}

\begin{figure}
\begin{center}
 \includegraphics[width=\linewidth]{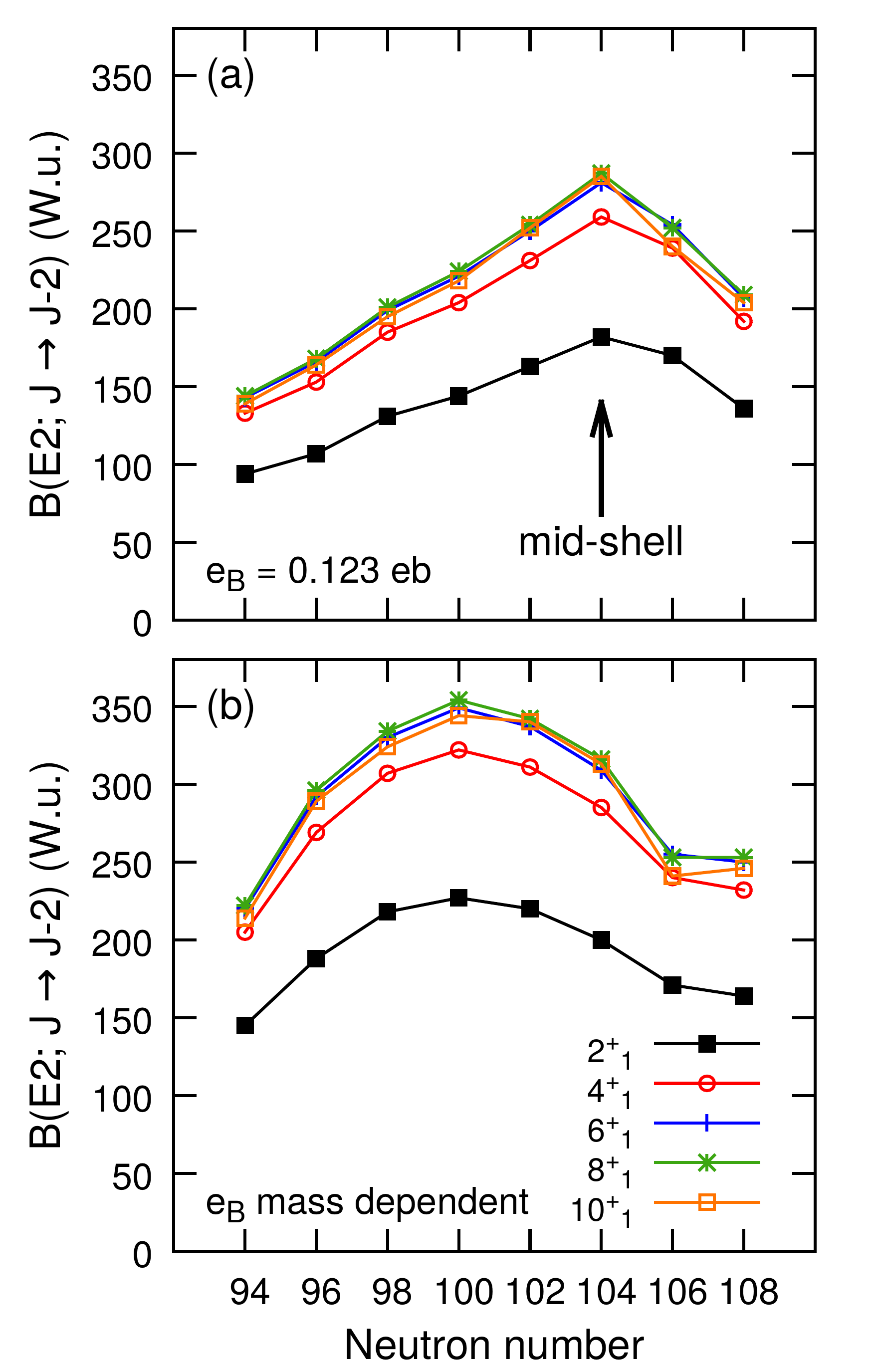}
\end{center}
\caption{(Color online) 
The $B$(E2) values for hafnium ($Z$=72) isotopes calculated with
two different choices of effective boson charge $e_\mathrm{B}$: 
(a) $e_\mathrm{B} = 0.123 \mathrm{eb}$ fixed for all isotopes; 
(b) $e_{B}$ different from nucleus to nucleus, each determined by
 adjusting the transition quadrupole moment $Q_t(2^+_1 \rightarrow 
0^+_1)$ from the IBM-2 calculation to fit $\beta_{2,
\mathrm{min}}$ of the Gogny-HFB D1M calculation. 
The boson effective charges $e_{B}$ used for the latter calculation are tabulated in
 Table~\ref{tab:IBMpara}. 
} 
\label{fig:theo_BE2}
\end{figure}

Let us now turn our attention to the calculation of the $B$(E2) values. 
In Fig.~\ref{fig:theo_BE2}(a) the theoretical $B$(E2;$J\rightarrow
J-2$) values ($J=2,4,\ldots, 10$) calculated with an effective charge 
$e_B=0.123$ $e$b are shown. The effective charge was fixed for all the
considered Hf nuclei as to reproduce
the experimental $B({\textnormal{E2}};2^+_1\rightarrow 0^+_1)$ value 
 of 182 W.u in $^{176}$Hf. 
As anticipated from  previous IBM fitting calculations for W and Os
isotopes \cite{Rudigier201089}, the theoretical $B$(E2;$J\rightarrow
J-2$) values with a fixed $e_B$ value become maximal at mid-shell
$N=104$, which disagrees with the experimental 
$B$(E2) systematics showing a peak at $N=100$ or 102.

The failure can be partly attributed to the use of a fixed effective
charge $e_\mathrm{B}$. 
Then, we consider $e_\mathrm{B}$ to be mass dependent and determine its
value for each individual nucleus. 
Specifically, we propose to derive $e_\mathrm{B}$ so that it follows the 
systematics of
the Gogny-HFB $\beta_{2,min}$ values, mainly because this quantity reaches its maximum value not at
mid-shell but at $N=100$, as seen from Fig.~\ref{fig:theo_beta}.  

A possible option to take this effect into account can be to associate the transition quadrupole moment
$Q_t$, corresponding to the $2^+_1\rightarrow 0^+_1$ E2 transition
matrix element, to the Gogny-HFB $\beta_{2,min}$ value. 
%
In general, $Q_{t}$ for the transition from the state with spin $J$ to
$J^{\prime}$ is written as
\begin{eqnarray}
 Q_t(J\rightarrow
  J^{\prime})=\sqrt{\frac{16\pi}{5}\frac{B({\textnormal{E2}};J\rightarrow
  J^{\prime})}{(J200|J^{\prime}0)^2}}, 
\end{eqnarray}
with $(J200|J^{\prime}0)$ being the Clebsch-Gordan coefficient. The quantity
$Q_t(J\rightarrow J^{\prime})$ is transformed into the intrinsic deformation parameter
$\beta_t(J\rightarrow J^{\prime})$
\begin{eqnarray}
 \beta_t(J\rightarrow
  J^{\prime})=\frac{\sqrt{5\pi}}{3ZR^2}Q_t(J\rightarrow J^{\prime}). 
\end{eqnarray}
The $\beta_t$ for the $2^+_1\rightarrow 0^+_1$ E2 transition,
$\beta_t(2^+_1\rightarrow 0^+_1)$, is eventually made equal, for each
individual nucleus, to the
Gogny-HFB mean-field $\beta_{2,min}$ in order to obtain the effective charge 
$e_\mathrm{B}$.

In Fig.~\ref{fig:theo_BE2}(b) we plot the resulting
$B$(E2;$J\rightarrow J-2$) values with the effective charge determined
in this way. 
Although the $B$(E2;$J\rightarrow J-2$) values in panel (b) are
generally larger in magnitude than those in panel (a), the systematic
trend is more consistent with the experimental one, as it becomes
maximal at $N=100$, not at the mid-shell nucleus $^{176}$Hf. 
Again, one notices that the result is well correlated with the Gogny-HFB $\beta_{2,min}$ value
(Fig.~\ref{fig:theo_beta}), the Thouless-Valatin MOI
(Fig.~\ref{fig:theo_moments}), and the variation of the extracted $e_\mathrm{B}$
value as a function of $N$ (Tab.~\ref{tab:IBMpara}).

The observed experimental $B$(E2) systematics in the neutron-deficient
Hf isotopes, indicating the maximum collectivity not at the middle of the
major shell, implies that for the analysis of realistic nuclei a simple
model may not be sufficient and that certain microscopic effects have to
be taken into account. 
In particular, the mass-dependence of $e_\mathrm{B}$ implies that the effect of
the core polarization may become non-negligible. 
In the present IBM framework, the polarization effect cannot be
included explicitly, but it could be somehow taken into account by absorbing it in the variation
of the boson charge.

Alternatively, the mass dependence of the
boson effective charge could be explained by the renormalization effect of the $g$
boson \cite{taka85}. 
In the $sd$ IBM system, the $g$-boson effect could show up as
higher-order terms in the quadrupole operator \cite{taka85}. 
In that sense, the form of the quadrupole operator used in the present
work in Eq.~(\ref{eq:bh}), which is a one-body operator, can be
extended to include the higher-order terms.

\begin{figure}
\begin{center}
 \includegraphics[width=\columnwidth]{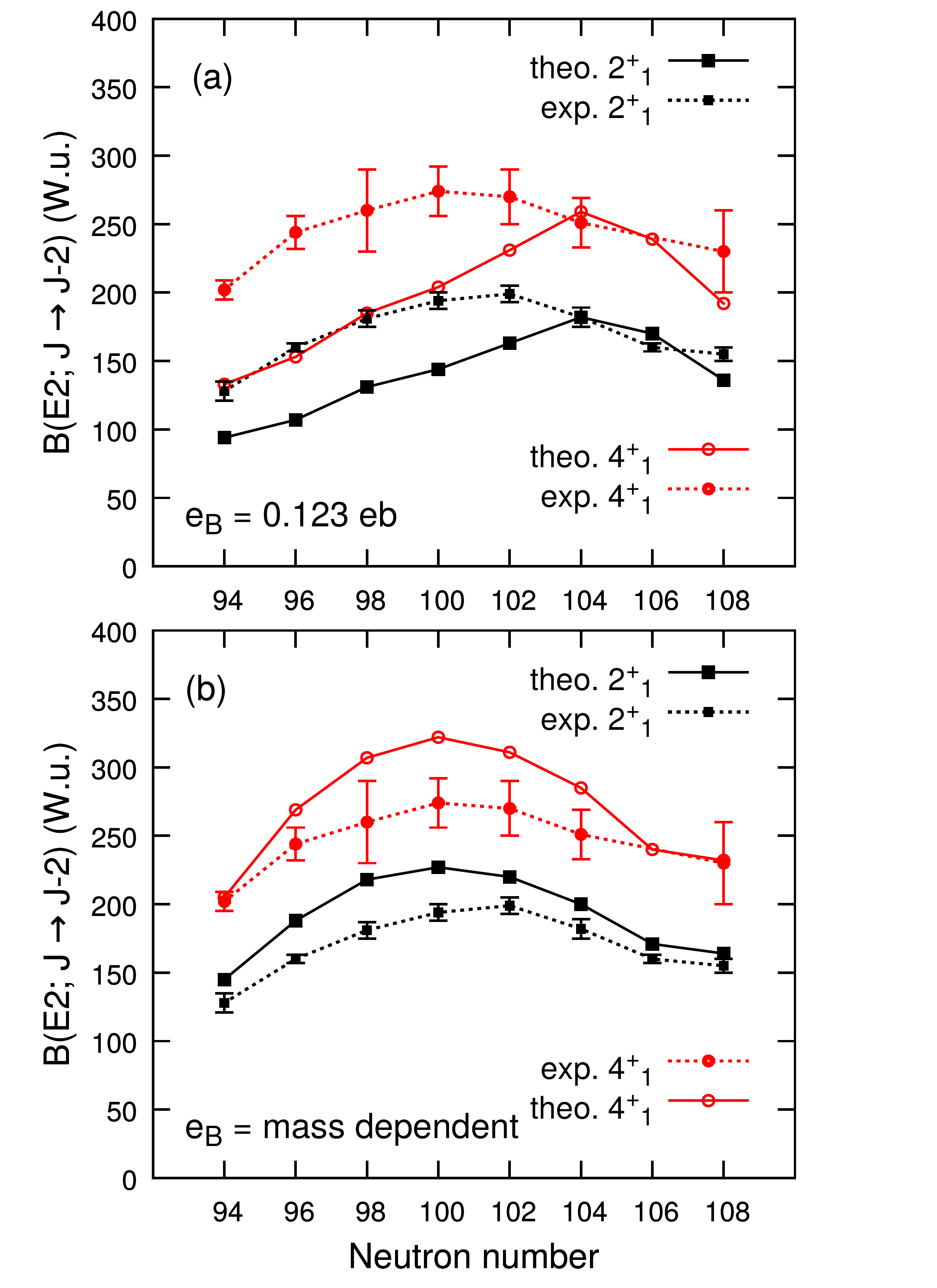}
\end{center}
\caption{(Color online) 
Comparison of experimental and theoretical reduced
transition strengths in hafnium isotopes (Z=72). (a) Calculation from Fig.
\ref{fig:theo_BE2} (a) ($e_\mathrm{B} = 0.123 \mathrm{eb}$). (b) Calculation 
from Fig. \ref{fig:theo_BE2} (b) ($e_\mathrm{B}$ mass dependent).
} 
\label{fig:theo_exp_BE2_2_4}
\end{figure}

We also comment on the dependence of the IBM results on the choice of
the EDF parametrization. 
For the energies of the ground-state rotational band, there is
a certain quantitative difference between the D1S and the D1M results,
but the overall tendency at the qualitative level is expected to be quite similar, because the
deformation energy surface, $\beta_{2,min}$ and Thouless-Valatin MOI for
both parametrizations have been shown to exhibit similar features as a function of neutron
number. 
Likewise, for the $B$(E2) systematics, as both the D1S and D1M sets provide a
similar trend of the $\beta_{2,min}$ value 
(cf. Fig.~\ref{fig:theo_beta}), if the effective charges are determined
in the same way, one would obtain the results
qualitatively similar to the ones from the D1M interactions
(cf. Figs.~\ref{fig:theo_BE2} and \ref{fig:theo_exp_BE2_2_4}). 

\begin{figure}
\begin{center}
 \includegraphics[width=\columnwidth]{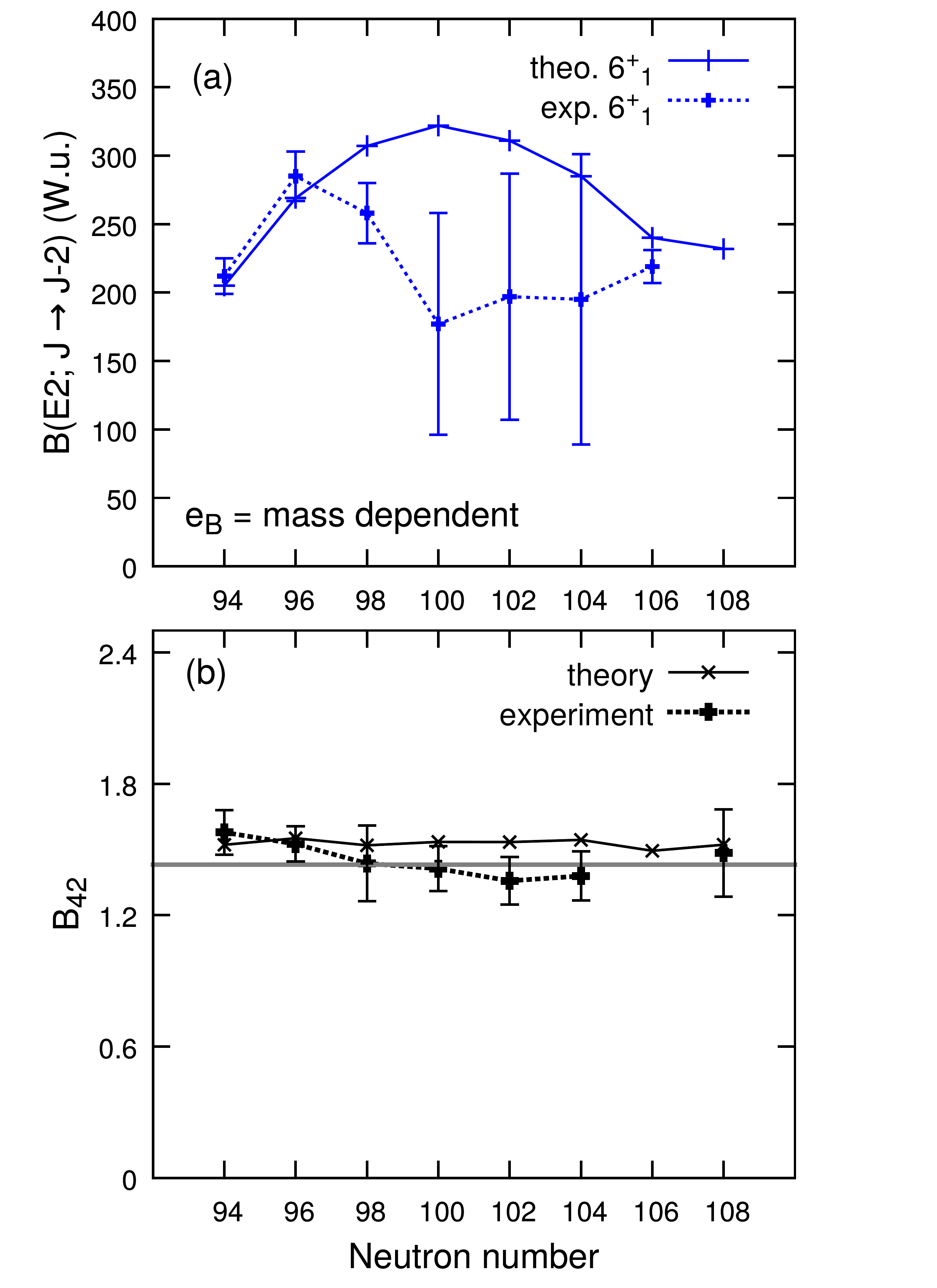}
\end{center}
\caption{(Color online) 
Comparison of experimental and theoretical reduced
transition strengths in hafnium isotopes (Z=72).  (a) Calculation
from Fig. \ref{fig:theo_BE2} (b) ($e_\mathrm{B}$ mass dependent). (b) $B_{42}$ 
as defined in the text. The IBM SU(3) limit of $B_{42}=10/7$ is indicated as a 
gray line.
} 
\label{fig:theo_exp_BE2_6_R}
\end{figure}

For completeness, in Fig.~\ref{fig:theo_exp_BE2_2_4} we compare the
$B$(E2;$4^+_1\rightarrow 2^+_1$) and $B$(E2;$2^+_1\rightarrow 0^+_1$) experimental
values with the results of the calculations with two different
choices of the effective charge: (a) $e_\mathrm{B}=0.123$ $e$b fixed for all 
nuclei, and (b) $e_{B}$ dependent on mass. 
In Fig.~\ref{fig:theo_exp_BE2_6_R}, we compare the experimental and theoretical (with
the use of mass-dependent effective charge) $B$(E2;$6^+_1\rightarrow 4^+_1$) value (a)
and the $B_{42}$ ratio (b). 
It is rather evident from Figs.~\ref{fig:theo_exp_BE2_2_4}(a,b) that the 
calculations with the fixed boson effective
charge do not reproduce the experimental trend while a nice agreement
between theory and calculation is obtained if one chooses
the effective charge for each nucleus as to follow the variation 
of the $\beta_{2,min}$ value with $N$. 
Overall, a reasonable description of the data is obtained both
qualitatively and quantitatively. 

In Fig.~\ref{fig:theo_exp_BE2_2_4}(b), the calculated $B$(E2;$4^+_1\rightarrow 
2^+_1$) and
$B$(E2;$2^+_1\rightarrow 0^+_1$) values overestimate the experimental
values for the nuclei around $N=100$. 
This might be due to the presence of a non-zero hexadecapole deformation
$\beta_4$. 
A non-zero $\beta_4$ has an influence on the transition quadrupole moment $Q_t$, and
therefore on the extracted effective charge. Likely, the inclusion of $\beta_4$ will 
improve agreement with experiment for these nuclei. 
In fact, the previous IBM-1 calculation indicated \cite{zerguine08} that the
hexadecapole degree of freedom may be non-negligible, though in a
different context of some ground-state properties.

For the $6^{+}_1\rightarrow 4^+_{1}$ E2 transition presented in
Fig.~\ref{fig:theo_exp_BE2_6_R}(a), the theoretical value 
shows a parabolic behavior. 
The experimental one does not show this trend, but also has a large
uncertainty.
In Fig.~\ref{fig:theo_exp_BE2_6_R}(b), we observe that both the experimental and
theoretical $B_{42}$ ratio are close to each other, as well as to the
SU(3) limit of the IBM (=10/7) \cite{IBM}, showing that the considered nuclei 
are
well described as being good rotors.

In comparison to the other EDF-based calculation available in the
literature, the five-dimensional collective Hamiltonian 
approach based on the Gogny-D1S parameter set has predicted the maximum
$B$(E2;$2^+_1\rightarrow 0^+_1$) value at $102$ \cite{CEA}. 
It is one unit different from our result, but in that calculation 
the difference in the $B$(E2;$2^+_1\rightarrow 0^+_1$) values 
between $^{172}$Hf and $^{174}$Hf is negligible.




\section{Conclusion\label{sec:conclusion}}

The experimental technique of fast timing using a combination of LaBr detectors
and a conversion electron spectrometer, is very well suited to measure yrast
B(E2) values in deformed nuclei.
It was possible to measure the half-lives of ground state band excitations of a
nucleus from the $2^+_1$ to the $6^+_1$ state in one single experiment with a
relatively short amount of measuring time. E.g. 48 hours of net measuring time
in the case of $^{172}$Hf. The GCD method for extracting the shorter half-lives
gave consistent results for different decay-feeder energy combinations. The
detection limit for picosecond half-lives, given a good peak-to-background 
ratio, was around 5 ps.

In the present work the half life of the $2^+_1$ state in the nuclei $^{172,174,176}$Hf were remeasured. 
The results are in agreement with those from Coulomb excitation measurements and other methods. The deviation with 
respect to results of $\gamma\gamma$ fast timing measurements from the late 1960s can possibly be explained by the 
improved energy resolution of the LaBr detectors and the use of the Orange spectrometer in the current work, both 
of which allow for a better treatment of background and contaminations.
Furthermore the half lives of the $4^+_1$ and $6^+_1$ state in the nuclei $^{172,174,176}$Hf were measured for the 
first time. It was possible to deduce an upper limit for the half life of the $8^+_1$ state in the same nuclei.

The evolution of $B(E2)$ transition strength in the considered even-even
hafnium isotopes is smooth, as was expected. The maximum of the $B(E2; 2^+_1
\rightarrow 0^+_1)$ value is found not at mid-shell $N=104$, but at lower 
neutron
number, which in this case has turned out to be $N=102$. 
A simple model cannot explain this systematics, only predicting the
maximum collectivity at the mid-shell. 
One should rather rely on a more realistic or microscopic model for
complex nuclei. 
Our spectroscopic calculation, performed within the
scope of this work, 
has reproduced the experimental trend of the $B$(E2) very nicely (see Fig.~\ref{fig:theo_BE2}(c)),
if the boson effective charge is determined as to follow the prediction
by the microscopic EDF calculation. 
In the present calculation, any specific adjustment to the data has not
been invoked, but the result depends only on the EDF parametrization
and on the mapping procedure. 
The results of microscopic EDF calculations on the quadrupole
deformation $\beta_{2,min}$ (Fig.~\ref{fig:theo_beta}) and the cranking
moment of inertia (Fig.~\ref{fig:theo_moments}), the corresponding IBM
energy levels (Fig.~\ref{fig:theo_energies}) and E2 transition rates
(Figs.~\ref{fig:theo_BE2} and \ref{fig:theo_exp_BE2_2_4}) are consistent and 
correlated with each other very well in systematics. 

On the other hand, the same behavior of the $B$(E2) transition strength
has also been observed in neighboring
isotopic chains like erbium, ytterbium and tungsten. 
In that sense, it would be a very interesting subject of a future study
to analyze the spectroscopy of these neighboring nuclei in a more
systematic manner.

\begin{acknowledgements}
This work has been supported by the DFG under grant JO 391/16-1. 
K. N. acknowledges the support by the Marie Curie Actions grant within
the Seventh Framework Program of the European Commission under Grant
No. PIEF- GA-2012-327398.  
This work has been partially carried out during his visit to the
 Institut f\"ur Kernphysik (IKP) of the University of Cologne. He
 acknowledges Prof. J. Jolie and the IKP for their warm hospitality. 
\end{acknowledgements}

\appendix

\section{Procedure to extract parameters for the IBM Hamiltonian\label{sec:mapping}}

The IBM-2 Hamiltonian of Eq.~(\ref{eq:bh}) contains 5 free parameters ($\epsilon$,
$\kappa$, $\chi_{\pi}$, $\chi_{\nu}$ and $\kappa^{\prime}$) to be
determined. 
The bosonic energy surface $E_{\textnormal{IBM}}(\beta,\gamma)$ is given by an analytical expression,
and is derived by taking the 
expectation value of the IBM-2 Hamiltonian in Eq.~(\ref{eq:bh}) with the
boson coherent state $|\phi(\beta,\gamma)\rangle$ \cite{GK}: 
\begin{eqnarray}
 |\phi(\beta,\gamma)\rangle=\Pi_{\tau}\frac{1}{\sqrt{N_{\tau}!}}(\lambda^{\dagger}_{\tau})^{N_{\tau}}|0\rangle
\end{eqnarray}
where
$N_{\tau}$ and $|0\rangle$ represent the number of proton or neutron
bosons and inert core, respectively. 
\begin{eqnarray}
 \lambda_{\tau}^{\dagger}=s_{\tau}^{\dagger}+\frac{1}{\sqrt{2}}\bar\beta_{\tau}\sin{\gamma_{\tau}}(d^{\dagger}_{\tau
  +2}+d^{\dagger}_{\tau
  -2})+\bar\beta_{\tau}\cos{\gamma_{\tau}}d^{\dagger}_{\tau} \nonumber \\
\end{eqnarray}
$\bar\beta_{\tau}$ is the deformation parameter in the boson system, and
we assume $\bar\beta_{\pi}=\bar\beta_{\nu}\equiv\bar\beta$. 
In the above equations, we distinguish the deformation parameter $\bar\beta$ of the boson
system  from the usual $\beta$ deformation
parameter in the collective model. 
The model space of the collective model
spans the entire nucleus, while only the valence nucleons are considered in
the IBM. 
Therefore, the deformation parameter for the IBM system $\bar\beta$ is
always larger than the one in the collective model $\beta$, and one can 
assume, to a good approximation, that $\bar\beta\propto\beta$ \cite{GK}. 
This involves an additional parameter $C_{\beta}$, which is the proportionality
coefficient for the $\beta$ deformation, $\bar\beta=C_{\beta}\beta$

The boson energy surface is given as 
\begin{eqnarray}
\label{eq:bpes}
E_{\textnormal{IBM}}(\beta,\gamma)
&=&\frac{\langle\phi(\beta,\gamma)|\hat
H^{\textnormal{IBM}}|\phi(\beta,\gamma)\rangle}{\langle\phi(\beta,\gamma)|\phi(\beta,\gamma)\rangle}
\nonumber \\
&=&\frac{\epsilon^{\prime}(N_{\pi}+N_{\nu})\bar\beta^2}{1+\bar\beta^2}
+
\frac{N_{\pi}N_{\nu}\kappa}{(1+\bar\beta^2)^2}\nonumber\\
&&\times\Big[
4\bar\beta^2-4\sqrt{\frac{2}{7}}(\chi_{\pi}+\chi_{\nu})\bar\beta^3\cos{3\gamma}\nonumber\\
&&+\frac{2}{7}\chi_{\pi}\chi_{\nu}\bar\beta^4\Big], 
\end{eqnarray}
with $\epsilon^{\prime}=\epsilon-6\kappa^{\prime}$ and $\bar\beta=C_{\beta}\beta$. 

The four parameters $\epsilon^{\prime}(=\epsilon-6\kappa^{\prime})$,
$\kappa$, $\chi_{\pi}$ and $\chi_{\nu}$ plus the additional coefficient
$C_{\beta}$ are determined by adjusting the bosonic energy surface
$E_{\textnormal{IBM}}(\beta,\gamma)$ so that it reproduces the topology of the microscopic energy surface
$E_{\textnormal{HFB}}(\beta,\gamma)$ in the neighborhood of the absolute
minimum. 
This procedure reduces to the fitting of the bosonic energy surface in
Eq.~(\ref{eq:bpes}) to the Gogny-D1M energy surface. 
For the fit, we utilize the technique using the wavelet transform \cite{Nom10}. 

On the other hand, the $\hat L\cdot\hat L$ term does not contribute to the energy
surface of Eq.~(\ref{eq:bpes}) but takes the same analytical expression
as the first term in Eq.~(\ref{eq:bh}). 
Therefore, once the above five parameters are obtained, the
coefficient $\kappa^{\prime}$ should be determined 
in a different way from the other five parameters. 
To do this, we take the procedure of Ref.~\cite{Nom11}: the cranking moment of
inertia is compared between fermion and boson systems. 
We then calculate the MOI for the $2^{+}_{1}$ state by
Thouless-Valatin (TV) formula \cite{TV}: 
\begin{eqnarray}
\label{eq:tv}
 {\cal I}_{\textnormal{TV}}=\frac{J(J+1)}{2E_{\gamma}}. 
\end{eqnarray}
$E_{\gamma}$ stands for the $2^{+}_{1}$ excitation energy obtained
from the self-consistent cranking calculation with the constraint
$\langle\hat J_{x}\rangle=\sqrt{J(J+1)}$, where $\hat J_x$ represents
the $x$ component of the angular momentum operator. 

The equivalent quantity is derived for the IBM in the coherent state
$|\phi(\beta,\gamma)\rangle$, using the cranking formula of Schaaser and
Brink \cite{Schaaser86}: 
\begin{eqnarray}
\label{eq:bmom}
 {\cal
  I}_{\textnormal{IBM}}=\lim_{\omega\rightarrow\infty}\frac{1}{\omega}\frac{\langle\phi(\beta,\gamma)|\hat
  L_{x}|\phi(\beta,\gamma)\rangle}{\langle\phi(\beta,\gamma)|\phi(\beta,\gamma)\rangle}, 
\end{eqnarray}
with $\omega$ being the cranking frequency. 

With the parameters $\epsilon^{\prime}(=\epsilon-6\kappa^{\prime})$,
$\kappa$, $\chi_{\pi}$, $\chi_{\nu}$ and $C_{\beta}$, fixed from the 
energy-surface fit, the IBM moment of
inertia in Eq.~(\ref{eq:bmom}) contains only one parameter
$\kappa^{\prime}$. 
The $\kappa^{\prime}$ value is determined so that ${\cal
I}_{\textnormal{IBM}}$ calculated at the equilibrium point, where the
energy surface is minimal, is equal to the ${\cal
I}_{\textnormal{TV}}$ value at the corresponding energy minimum.

\bibliography{Hf_BE2s}

\end{document}